\documentclass[a4paper,11pt,twoside]{article}
\usepackage{mathrsfs}
 \usepackage[utf8]{inputenc}
 \usepackage{epsfig,graphicx,color}
 \usepackage{subfig}
 \usepackage{latexsym}                 % extension symboles
 \usepackage{amsmath,amsfonts,verbatim,xkeyval,bm, amsthm, upgreek, amscd, wasysym, amssymb, amsbsy}
 \usepackage{cases}
\usepackage{bm}
 \usepackage[english]{babel}
 \usepackage{listings}
 \usepackage{multicol}
 \usepackage[colorlinks]{hyperref}  %serve per fare collegamenti a pagine web
 \hypersetup{hidelinks}  %serve per fare i collegamenti in nero
 \usepackage{tikz-cd}
 \usepackage{booktabs}
\usepackage[numbers, sort&compress]{natbib}
 \usepackage{sidecap}
 \usepackage[font=small,labelfont=bf]{caption}
 \usepackage{float}
\usepackage{titlesec}
\usepackage{makecell}
\usepackage{soul}
\usepackage{xcolor,colortbl}
\usetikzlibrary{positioning}

\titleformat{\paragraph}
{\normalfont\normalsize\bfseries}{\theparagraph}{1em}{}
\titlespacing*{\paragraph}
{0pt}{3.25ex plus 1ex minus .2ex}{1.5ex plus .2ex}
 
\DeclareMathAlphabet\mathbfcal{OMS}{cmsy}{b}{n}

\def\avg#1{\left\langle #1 \right\rangle}

\def\nNewt{{n_{{\cal N}}}} 
\def\vet#1{{\bm #1}}
\def\reali{\mathbb{R}}
\def\complessi{\mathbb{C}}
\def\naturali{\mathbb{N}}
\def\interi{\mathbb{Z}}
\def\toro{\mathbb{T}}

\def\e{\mathrm{e}}
\def\i{\mathrm{i}}

\def\C{\mathscr{C}}

\def\F{\mathscr{F}}
\def\Ascr{\mathcal{A}}

\def\Escr{\mathcal{E}}

\def\Hscr{\mathcal{H}}

\def\Kscr{\mathcal{K}}

\def\Oscr{\mathcal{O}}

\def\Sscr{\mathcal{S}}

\def\Kscr{\mathcal{K}}

\def\b{\mathbf{b}}
\def\c{\mathbf{c}}
\def\d{\mathbf{d}}
\def\Ups{$\upsilon$-Andromed{\ae}~}
\def\ups{$\upsilon$-And~}
\def\rho{\varrho}
\def\csi{\xi}

\def\poisson#1#2{\lbrace #1,#2 \rbrace}

\def\t#1{\tilde{#1}}

\definecolor{lavender}{rgb}{0.9, 0.9, 0.98}

\newtheorem{theorem}{Theorem}[section]
\newtheorem{theorem*}{Theorem}
\newtheorem{lemma}{Lemma}[section]

\newcommand\blfootnote[1]{%
  \begingroup
  \renewcommand\thefootnote{}\footnote{#1}%
  \addtocounter{footnote}{-1}%
  \endgroup
}

\title{\bf Computer-assisted proofs of existence of KAM
  tori\\ in planetary dynamical models of \ups$\b$}
\author{
  {\bf Rita Mastroianni}\\
  {\small Dipartimento di Matematica ``Tullio Levi-Civita'', Universit\`a degli Studi di Padova,}\\
  {\small via Trieste 63, 35121 Padova,}\\
  {\bf Ugo Locatelli}\\
  {\small Dipartimento di Matematica, Universit\`a
  degli Studi di Roma ``Tor Vergata'',}\\
  {\small via della Ricerca Scientifica 1, 00133 Roma,}\\
  {\small e-mails:
  {\tt rita.mastroianni@math.unipd.it, locatell@mat.uniroma2.it}}
}

\date{}

\begin{document}
\maketitle

\markboth{R. Mastroianni, U. Locatelli}{Computer-assisted proofs
  $\ldots$ in planetary dynamical models of \ups$\b$}


\begin{abstract}
  \noindent
  We reconsider the problem of the orbital dynamics of the innermost
  exoplanet of the $\upsilon$-Andromed{\ae} system (i.e., \ups$\b$)
  into the framework of a Secular Quasi-Periodic Restricted
  Hamiltonian model. This means that we preassign the orbits of the
  planets that are expected to be the biggest ones in that extrasolar
  system (namely, \ups$\c$ and \ups$\d$). The Fourier decompositions
  of their secular motions are injected in the equations describing
  the orbital dynamics of \ups$\b$ under the gravitational effects
  exerted by those two exoplanets. By a computer-assisted procedure,
  we prove the existence of KAM tori corresponding to orbital motions
  that we consider to be very robust configurations, according to
    the analysis and the numerical explorations made in our previous
    article.  The computer-assisted assisted proofs are successfully
  performed for two variants of the Secular Quasi-Periodic Restricted
  Hamiltonian model, which differs for what concerns the effects of
  the relativistic corrections on the orbital motion of \ups$\b$,
  depending on whether they are considered or not.
\end{abstract}

%%%%%%%%%%%%%%%%%%%%%%%%%%%%%%%%%%%%%%%%%%%%%%%%%%%%%%%%%%%%%%%%%%%%%%%%%
\section{Introduction}
\label{sec:introduction}
The longstanding tradition of the study of the planetary orbital
dynamics is mainly due to the fact that it is a fascinating
problem. However, it also constitutes a prototypical example of
system sharing some fundamental features with other models that are of
great interest in physics.  We limit ourselves to mention two of these
properties characterizing also other fundamental dynamical
problems. First, in the planetary models different time-scales of
evolution are often easy to recognize: a fast dynamics, corresponding
to the periods of orbital revolution and a {\it secular} one, which
describes the slow deformation of the Keplerian orbits (see,
e.g.,~\cite{las1996} for an introduction including also some
historical notes). Moreover, it is natural to consider systems with
planets much bigger than others into the framework of {\it
  hierarchical} models. In such a situation, the dynamics of the
subsystem including the major bodies can be studied separately. More
precisely, it is possible to prescribe, first, the motion of such
major bodies, by preliminarly determining a solution (that can be
analytical or numerical) of the subsystem hosting the near totality of
the mass, and then devote the focus on the orbital evolution of the
smaller bodies under the gravitational attraction exerted by the
larger ones. In the last decade, this approach allowed to obtain
interesting results about the source of the long-term instability of
the terrestrial planets of our Solar System (see,
e.g.,~\cite{batetal2015} and~\cite{hoaetal2022}).  In our previous
work (see~\cite{masloc2023}), the problem of the long-term stability
of the possible orbital configurations of \ups$\b$ was numerically
studied in the framework of a model that is both {\it secular} and
{\it hierarchical}.

Four exoplanets orbiting around \Ups{A}, that is the brightest star of
a binary hosting also the red dwarf \Ups{B}, have been detected so far
(see~\cite{butetal1999} and~\cite{curetal2011}). Such exoplanets are
named \ups$\b$, $\c$, $\d$ and ${\bf e}$ in alphabetic order going out
from the main star. Since \Ups{B} is very far with respect to these
other bodies (i.e., $\sim 750\,$AU), then it is usual to neglect its
gravitational effects when the planetary system orbiting around
\Ups{A} is studied. Let us also recall that none of the currently
available observational methods allows us to know all the dynamical
parameters characterizing an extrasolar planet. In particular, the
technique used to detect these four exoplanets (i.e., the so called
Radial Velocity method) provides just the minimal possible value of
their masses.  However, in the case of \ups$\c$ and \ups$\d$, which
are expected to be the biggest planets in that extrasolar system, some
additional data taken from the Hubble Space Telescope allowed to give
a more accurate evaluation of their masses, although with remarkable
uncertainties (see~\cite{mcaetal2010}). Moreover, the minimal mass of
\ups{\bf e} is one order of magnitude smaller than the one of
\ups$\d$, therefore, its effects on the orbital dynamics of the
innermost exoplanet of the system can be considered negligible.  The
question of the orbital stability of the three exoplanets closest to
\Ups{A} is quite challenging, since numerical integrations revealed
that unstable motions are frequent (see~\cite{deietal2015}). In order
to study and to analyze such a kind of model, we propose the following
strategy. First, by using a numerical criterion inspired from normal
form theory and introduced in~\cite{locetal2021}, we select the most
robust orbital configuration of the subsystem including the exoplanets
expected to be the biggest ones, that are \ups$\c$ and \ups$\d$. Such
a configuration has to be considered quite probable because it has to
persist under the perturbations exerted by both the innermost
exoplanet, i.e., \ups$\b$, and the outermost one, namely,
\ups$\bf{e}$. Furthermore, the average distance of \ups$\b$ from the
star is about $1/12$ of the one of \ups$\c$, which is, in turn, $1/3$
with respect to that concerning \ups$\d$.  Moreover, the minimal mass
of \ups$\b$ is one order of magnitude smaller than the ones of \ups
$\c$ and \ups $\d$, which are expected to be the largest
ones. Therefore, it is very natural to assume that the mutual
interaction between \ups$\c$ and \ups $\d$ is much more relevant with
respect to those with the innermost exoplanet. For all these reasons,
in~\cite{masloc2023}, we modeled the long-term evolution of \ups$\b$
introducing a restricted four-body problem. In more details, the
secular motions of the major exoplanets \ups $\c$ and \ups $\d$
(corresponding to the quasi-periodic orbit that is expected to be the
most robust) were approximated by the truncated Fourier expansions
provided by the well known technique of the Frequency Analysis (see,
e.g.,~\cite{las2005}). These quasi-periodic laws of motion were
injected in the equations describing the orbital dynamics of
\ups$\b$. Such a procedure allowed us to drastically simplify the
problem, because we introduced a \textit{Secular Quasi-Periodic
  Restricted} (hereafter, SQPR) Hamiltonian model having $2+3/2$
degrees of freedom instead of $9$, as it was for the original complete
four-body problem, after the reduction of the center of
mass. In~\cite{masloc2023}, such a simplified model was analyzed in a
mainly numerical way, with the aim to describe the regions of
dynamical stability of \ups $\b$, as a function of the unknown initial
values of the obital parameters, i.e., the longitudes of the node
$\Omega$ and the inclination $\i$.  In the present work this stability
problem is reconsidered in a less extensive way, but with more
mathematical purposes, in order to prove that there exist KAM tori
which include a few possible particular orbits of \ups$\b$ that are in
agreement with the currently available observations.

At least in a local sense, extremely strict upper bounds on the
diffusion rate can be provided in the vicinity of invariant tori
(see~\cite{morgio1995} for the explanation of the general strategy
and~\cite{giolocsan2017} for an application to a planetary system).
Therefore, the proof of the KAM theorem can also be seen as a
cornerstone in the design of a strategy which aims at reaching results
on the long-term stability of a Hamiltonian system with more than two
Degrees Of Freedom (hereafter, often replaced by DOF). Usually, purely
analytical proofs of existence of KAM tori are not enough to provide
results of interest for the study of realistic physical models;
therefore, in this context, a few different approaches have been
developed in order to produce Computer-Assisted Proofs (hereafter,
often replaced by CAPs). For instance, in the last decades the CAPs
were successful in ensuring the stability of a few relevant models of
planetary orbital dynamics with two DOF; this was done by applying a
topological confinement argument after having preliminarly proved the
existence of KAM tori acting as barriers for the motion (see,
e.g.,~\cite{locgio2000} and~\cite{celchi2007}). CAPs based on the so
called ``a-posteriori'' approach (see, e.g.,~\cite{calceldel2022} and
references therein) are able to prove the existence of KAM tori for
values of the small parameters very close to their breakdown threshold
in the simple and challenging framework of symplectic mappings
(see~\cite{figharluq2017}).  In~\cite{caletal2022}, this
``a-posteriori'' approach is adopted to give evidence of the existence
of KAM invariant tori for an interesting dissipative model in
Celestial Mechanics; the procedure is computer--assisted although it
does not provide a complete proof of such a result. In the present
work, we reconsider the problem of giving rigorous and complete CAPs
of existence of KAM tori for a couple of variants of the same
planetary Hamiltonian model, i.e., considering or not the effects of
the relativistic corrections on the orbital motion of \ups$\b$. Our
approach is based on a normal form method which recently succeeded in
proving the existence of KAM tori that are invariant with respect to
the slow dynamics of exoplanets whose orbits are trapped in some
resonances (see~\cite{caretal2022} and~\cite{danetal2023}). In this
paper, we further extend the range of the results obtained by using
this CAP method (that constructs Kolmogorov normal forms), in such a
way to be applied also to the SQPR Hamiltonian model describing the
orbital dynamics of the exoplanet \ups$\b$.  We emphasize that, as far
as we know, in the context of KAM theory this is the first complete
application of a CAP to a so realistic Hamiltonian model with more
than 2~DOF. We also stress that most of the technical work that is
needed in order to perform our CAP is deferred to a
code\footnote{``{\tt CAP4KAM{\_}nDOF}: Computer-Assisted Proofs of
  existence of KAM tori in Hamiltonian systems with $n$ ($\geq 2$)
  Degrees Of Freedom'', {\it Mendeley Data}, \url{
 https://doi.org/10.17632/tsffjx7pyr.2} (2022). Another version
    of that same software package, which includes also everything is
    needed in order to reproduce the CAPs discussed in
    Sections~\ref{sec:CAP-1stResult}--\ref{sec:CAP_REL} of the present
    paper, is available at the webpage
    \url{https://www.mat.uniroma2.it/\~locatell/CAPs/CAP4KAM-UpsAndB.zip} \label{footnote-CAP4KAM_nDOF}},
which can be downloaded from a public website. It is not uncommon to
allow any interested user to have access to the main core of a CAP;
for instance, the {\tt CAPD::DynSys} library is freely available and
was used to produce many Computer-Assisted Proofs for several
dynamical systems (see~\cite{kapetal2021} for an introduction
and~\cite{kapetal2022} for an example of application to the rigorous
computation of Poincar\'e maps). In our opinion, using the code
performing the CAPs as a ``black-box'', allows to really focus on the
main points of their strategy. In particular, a proof of existence of
invariant tori based on a normal form method essentially follows a
perturbative scheme. Therefore, most of the effort has to be devoted
to the search of a good enough approximation of the wanted
quasi-periodic solution, which will constitute the starting point for
the launch of the code performing the CAP. This also explains why we
focused nearly all the description of our work on the determination of
such an approximation of the final Kolmogorov normal form.

The present paper is organized as follows. In Section~\ref{sec:SQPR}
we recall the Secular Quasi-Periodic Restricted Hamiltonian describing
the dynamics of \ups $\b$, subject to the gravitational effects of the
exoplanets \ups $\c$ and $\d$. The application of the normal form
algorithm constructing elliptic tori to the SQPR model is described in
Section~\ref{sec:TE}. Section~\ref{sec:KAM} summarizes the reduction
allowing to pass to a $2+2/2$ DOF Hamiltonian model, the description
of the Kolmogorov normalization algorithm and its application (in
junction with a Newton-like method) to the $2+2/2$ SQPR Hamiltonian
model. The main result of the present work is presented in
Section~\ref{sec:CAP-1stResult}, where it is shown how the
computational procedure above can be performed preliminarly in order
to locate an approximation of the preselected orbit, which is a
good enough starting point to successfully complete the CAP of
existence of the wanted KAM torus.  All this computational procedure
is repeated in Section~\ref{sec:CAP_REL}, starting from a version of
the Secular Quasi-Periodic Restricted Hamiltonian model which includes
also relativistic corrections; this allows us to appreciate the
effects on the orbital dynamics due to General Relativity. Also in
this case, at the end of the Section, it is shown how the CAP is
successfully completed.

\section{The secular quasi-periodic restricted (SQPR) Hamiltonian model}
\label{sec:SQPR}

As explained in Section~$2$ of~\cite{masloc2023}, we need first to
prescribe the orbits of the giant planets \ups$\c$ and \ups$\d$. Thus,
we start from the Hamiltonian of the three-body problem (hereafter,
$3$BP) in Poincar\'e heliocentric canonical variables, using the
formulation based on the reduced masses $\beta_2\,$, $\beta_3\,$, that
is
\begin{align}
\label{Ham.3BP.reduced.mass}
\Hscr=&\sum_{j=2}^{3}\left( \frac{\vet{p}_j\cdot\vet{p}_j}{2\, \beta_j}
-\frac{\mathcal{G}\, m_0\, m_j}{r_j}\right)
+\frac{\vet{p_2}\cdot\vet{p_3}}{ m_0}
-\frac{\mathcal{G}\, m_2\, m_3}{|\vet{r_2}-\vet{r_3}|}\, ,
\end{align}
where $m_0$ is the mass of the star, $m_{j}\,$, $\vet{r}_j\,$,
$\vet{p}_j\,$, $j=2,3$, are the masses, astrocentric position vectors
and conjugated momenta of the planets, respectively, while $\mathcal{G}$ is
the gravitational constant and $\beta_j=m_0 m_j/(m_0+m_j)$, $j=2,3$,
are the reduced masses.\footnote{Let us remark that, in the following,
  we use the indexes $2$ and $3$ respectively, for the inner
  (\ups$\c$) and outer (\ups$\d$) planets between the giant ones,
  while the index $1$ is used to refer to \ups$\b$.}  Taking as
initial orbital parameters for the outer planets those reported in
Table~\ref{tab:param.orb} (and computing their corresponding values in
the Laplace reference frame, i.e., the invariant reference frame
orthogonal to the total angular momentum vector
$\vet{r}_2\times\vet{p}_2+\vet{r}_3\times\vet{p}_3\,$), we numerically
integrate the complete Hamiltonian~\eqref{Ham.3BP.reduced.mass} using
a symplectic method of type $\mathcal{SBAB}_3\,$
(see~\cite{lasrob2001}).

\begin{table}[h]
\begin{minipage}{0.5\textwidth}
\begin{center}
\begin{tabular}{lll}
\toprule
 & \ups$\c$ & \ups$\d$ \\
\midrule
$m \, [M_J]$ & $15.9792$ & $9.9578$ \\
$a(0) \, [{\rm AU}]$ & $0.829$ & $2.53$ \\
$\e(0)$ & $0.239$ & $0.31$ \\
$\i(0)\, [^{\circ}]$ & $6.865$ & $25.074$ \\
$M(0)\, [^{\circ}]$ & $355$ & $335$ \\
$\omega(0)\, [^{\circ}]$ & $245.809$ & $254.302$ \\
$\Omega(0)\, [^{\circ}]$ & $229.325$ & $7.374$ \\
\bottomrule
\end{tabular}
\end{center}
\end{minipage}
\begin{minipage}{0.4\textwidth}
\caption{Values of the masses and of the initial orbital parameters
  for \ups$\c$ and \ups$\d$, which are expected to correspond to
  their most robust orbital configuration, according to the numerical
  criterion discussed in~\cite{locetal2021}. These values are
  compatible with the observed data available, as reported
  in~\cite{mcaetal2010}, and the corresponding initial conditions in
  terms of astrocentric positions and conjugated momenta can be easily
  determined (see, e.g.,~\cite{murder1999}).}
\label{tab:param.orb}
\end{minipage}
\end{table}

The results produced by the numerical integrations can be expressed
with respect to the secular canonical Poincaré variables $(\csi_j,
\eta_j)$, $(P_j, Q_j)$ (momenta-coordinates) given by
\begin{equation}
\label{Poinc.var.U}
\begin{aligned}
  &\csi_j=\sqrt{2\Gamma_j}\cos (\gamma_j)=
  \sqrt{2\Lambda_j}\,\sqrt{1-\sqrt{1-\e_j^2}}\cos (\varpi_j)\, , \\
  &\eta_j=\sqrt{2\Gamma_j}\sin (\gamma_j)=
  -\sqrt{2\Lambda_j}\,\sqrt{1-\sqrt{1-\e_j^2}}\sin (\varpi_j)\, ,
  \qquad j=1,\,2,\,3\, ,\quad\\
  &P_j=\sqrt{2\Theta_j}\cos (\theta_j)=
  2\sqrt{\Lambda_j}\,\sqrt[4]{ 1-\e_j^2}\,
  \sin \left(\frac{\i_j}{2}\right)\cos( \Omega_j)\, ,\\
  &Q_j=\sqrt{2\Theta_j}\sin (\theta_j)=
  -2\sqrt{\Lambda_j}\,\sqrt[4]{ 1-\e_j^2}\,
  \sin \left(\frac{\i_j}{2}\right)\sin( \Omega_j)\, 
\end{aligned}
\end{equation}
where $\Lambda_j=\beta_j\sqrt{\mu_j a_j}\,$, $\beta_j=m_0
m_j/(m_0+m_j)\,$, $\mu_j=\mathcal{G}\left(m_0+m_{j}\right)\,$, and
$\e_j\,$, $\i_j\,$, $\omega_j\,$, $\Omega_j$,
$\varpi_j=\omega_j+\Omega_j$ refer, respectively, to the eccentricity,
inclination, argument of the periastron, longitudes of the node and of
the periastron of the $j$-th planet.

A good reconstruction of the motion laws $t\mapsto(\csi_j(t),
\eta_j(t))$, $t\mapsto(P_j(t), Q_j(t))$ ($j=2,3$), can be obtained by
the computational method of \textit{Frequency Analysis} (hereafter,
FA; see, e.g.,~\cite{las2005} ). Then, we use the FA to compute a
quasi-periodic approximation of the secular dynamics of the giant
planets \ups$\c$ and \ups$\d$, i.e., we determine two finite sets of
harmonics with the corresponding Fourier coefficients
$$
\Big\{ \big({k}_{j,s}\,,\,A_{j,s}\,,\,\vartheta_{j,s}\big)\in
       \interi^3\times\reali_+\times[0,2\pi)
       \Big\}_{{j=2,3\phantom{1234}}\atop{s=1,\ldots,\mathcal{N}_C}}
\ \ {\rm and} \ \          
\Big\{ \big(\tilde{k}_{j,s}\,,\,\tilde{A}_{j,s}\,,\,\tilde{\vartheta}_{j,s}\big)\in
       \interi^3\times\reali_+\times[0,2\pi)
       \Big\}_{{j=2,3\phantom{1234}}\atop{s=1,\ldots,\tilde{\mathcal{N}}_C}}
$$
such that       
\begin{equation}
\label{segnalicd}
\begin{aligned}
  &\csi_j(t)+i\eta_j(t)\simeq
  \sum_{s=1}^{\mathcal{N}_C}A_{j,s}e^{i(\vet{k}_{j,s}\cdot\vet{\theta}(t)+\vartheta_{j,s})}\, , \\ 
  &P_j(t)+iQ_j(t)\simeq
  \sum_{s=1}^{\widetilde{\mathcal{N}}_C}\tilde{A}_{j,s}
  e^{i(\tilde{\vet{k}}_{j,s}\cdot\vet{\theta}(t)+\tilde{\vartheta}_{j,s})}\, , 
\end{aligned}
\end{equation}
$\forall\ j=2,3\,$, where the angular vector 
\begin{equation}
\label{def_angle_theta_FA}
\vet{\theta}(t)=
(\theta_3(t), \theta_4(t), \theta_5(t))=(\omega_3\,t, \omega_4\, t, \omega_5\, t)
:=\vet{\omega}\,t
\end{equation}
depends \textit{linearly on time} and $\vet{\omega}\in\mathbb{R}^3$ is
the fundamental angular velocity vector whose components are listed in
the following:
%\vskip -.4truecm
\begin{align}
\begin{aligned}
\label{freq.fond.CD}
&\omega_3=-2.43699358194622660\times 10^{-3}\ {\rm rad/yr}, \\
&\omega_4=-1.04278712796661375\times 10^{-3}\ {\rm rad/yr}, \\
&\omega_5=\phantom{-}4.88477275490260560\times 10^{-3}\ {\rm rad/yr}.
\end{aligned}
\end{align}
Hereafter, the secular motion of the outer planets $t\mapsto
(\csi_j(t), \eta_j(t), P_j(t), Q_j(t))$, $j=2,3$, is approximated as
it is written in both the r.h.s. of formula~\eqref{segnalicd}. The
numerical values of 18 Fourier components (because ${\mathcal{N}}_C=4$
and $\tilde{\mathcal{N}}_C=5$) which appear in the quasi-periodic
decompositions of the motions laws as reported in Tables $2$, $3$,
$4$, $5$ of~\cite{masloc2023}.

Having preassigned the motion of the two outer planets \ups$\c$ and
\ups$\d\,$, it is now possible to properly define the secular model
for a quasi-periodic restricted four-body problem (hereafter,
$4$BP). We start from the Hamiltonian of the $4$BP, given by
\begin{equation}
\label{Ham.4BP.reduced.mass}
\Hscr_{4BP}=
\sum_{j=1}^{3}\left( \frac{\vet{p}_j\cdot\vet{p}_j}{2\, \beta_j}
-\frac{\mathcal{G}\, m_0\, m_j}{r_j}\right)
+\sum_{1\leq  i < j \leq 3}\frac{\vet{p_i}\cdot\vet{p_j}}{ m_0}
-\sum_{1\leq i < j \leq 3}\frac{\mathcal{G}\, m_i\, m_j}{|\vet{r_i}-\vet{r_j}|}\, .
\end{equation}
We recall that the so called secular model of order one in the masses
is given by averaging with respect to the mean motion angles, i.e.,
\begin{equation}
\label{average_Ham4BP}
\Hscr_{sec}(\vet{\csi}, \vet{\eta}, \vet{P}, \vet{Q})=
\>\int_{\toro^3}\!\!\!\!
\frac{\Hscr_{4BP}(\vet{\Lambda}, \vet{\lambda},
  \vet{\csi}, \vet{\eta}, \vet{P}, \vet{Q})}{8\pi^3}\,
{\rm d} \lambda_1 {\rm d}\lambda_2 {\rm d}\lambda_3\, .
\end{equation}
Due to the d'Alembert rules (see, e.g.,~\cite{murder1999}
and~\cite{mor2002}), it is well known that the secular Hamiltonian can
be expanded in the following way:
\begin{equation}
\label{H4BP_expl}
\Hscr_{sec}(\vet{\csi}, \vet{\eta}, \vet{P}, \vet{Q})=
\sum_{s=0}^{\mathcal{N}/2}\sum_{\substack{|\vet{i}|+|\vet{l}|+\quad\,\:\:\\
    |\vet{m}|+|\vet{n}|=2s}}\!\!\!\!\!\!
c_{\vet{i},\vet{l},\vet{m},\vet{n}}
\prod_{j=1}^{3}\csi_j^{i_j}\eta_j^{l_j} P_{j}^{m_j} Q_j^{n_j}\, ,
\end{equation}
where hereafter $|\vet{v}|$ denotes the $l_1$--norm of any (real or
integer) vector $\vet{v}$ and $\mathcal{N}$ is the order of
truncation in powers of eccentricity and inclination. We fix
$\mathcal{N}=8$ in all our computations.

In particular, in order to describe the dynamical secular evolution of
the innermost planet \ups$\b\,$, it is sufficient to consider the
interactions between the two pairs \ups$\b$, \ups$\c$ and \ups$\b$,
\ups$\d$. In more details, let
\begin{equation}
\label{Ham_3BP_scissor}
\Hscr^{\mathfrak{i}-\mathfrak{j}}_{sec}(\csi_{\mathfrak{i}}, \eta_{\mathfrak{i}},
P_{\mathfrak{i}}, Q_{\mathfrak{i}},\csi_{\mathfrak{j}},
\eta_{\mathfrak{j}}, P_{\mathfrak{j}}, Q_{\mathfrak{j}})=
\sum_{s=0}^{\mathcal{N}/2}\sum_{\substack{|\vet{i}|+|\vet{l}|+\quad\,\:\:\\
    |\vet{m}|+|\vet{n}|=2s}}\!\!\!\!\!\!
c_{\vet{i},\vet{l},\vet{m},\vet{n}}\prod_{j=\mathfrak{i},\mathfrak{j}}
\csi_j^{i_j}\eta_j^{l_j} P_{j}^{m_j} Q_j^{n_j}\, 
\end{equation}
be the secular Hamiltonian derived from the three-body problem for the
planets $\mathfrak{i}$ and $\mathfrak{j}$ (averaging with respect to
the mean longitudes $\lambda_\mathfrak{i}\,$,
$\lambda_\mathfrak{j}\,$).  We can finally introduce the
quasi-periodic restricted Hamiltonian model for the secular dynamics
of \ups$\b\,$; it is given by the following $2+3/2$ degrees of freedom
Hamiltonian:
\begin{equation}
\label{Ham.b.new}
\vcenter{\openup1\jot\halign{
 \hbox {\hfil $\displaystyle {#}$}
&\hbox {$\displaystyle {#}$\hfil}\cr
 \Hscr_{sec,\, 2+\frac{3}{2}}&(\vet{p},\vet{q},\csi_1, \eta_1, P_1, Q_1 )=
 \omega_3\,p_3 + \omega_4\,p_4 + \omega_5\,p_5
 \cr
 &\qquad+\Hscr^{1-2}_{sec}(\csi_1, \eta_1, P_1, Q_1, \csi_2(\vet{q}),
 \eta_2(\vet{q}), P_2(\vet{q}), Q_2(\vet{q}))
 \cr
 &\qquad+\Hscr^{1-3}_{sec}(\csi_1, \eta_1, P_1, Q_1,\csi_3(\vet{q}),
 \eta_3(\vet{q}), P_3(\vet{q}), Q_3(\vet{q}))\, ,
 \cr
}}
\end{equation}
where the pairs of canonical coordinates referring to the planets
\ups$\c$ and \ups$\d$ (that are $\csi_2\,$, $\eta_2\,$, \ldots
$P_3\,$, $Q_3$) are replaced by the corresponding finite Fourier
decomposition written in formula~\eqref{segnalicd} as a function of
the angles $\vet{\theta}$, renamed as $\vet{q}$ according to the notation
usually adopted in KAM theory, i.e.,
\begin{equation}
\label{rename_theta}
\vet{q}=(q_3\,,\,q_4\,,\,q_5):=
(\theta_3\,,\,\theta_4\,,\,\theta_5)=\vet{\theta}\, ,
\end{equation}
 $\vet{\omega}=(\omega_3,\omega_4,\omega_5)$ is the fundamental
angular velocity vector (defined in formula~\eqref{freq.fond.CD}) and
$\vet{p}=(p_3, p_4, p_5)$ is made by three so called ``dummy
actions'', which are conjugated to the angles $\vet{q}$. The equations
of motion for the innermost planet, in the framework of the restricted
quasi-periodic secular approximation, are described by
\begin{equation}
\label{campo.Ham.b}
\begin{cases}
  \dot{q_3}=\partial \Hscr_{sec,\,2+\frac{3}{2}}/\partial p_3=\omega_3\\
  \dot{q_4}=\partial \Hscr_{sec,\,2+\frac{3}{2}}/\partial p_4=\omega_4\\
  \dot{q_5}=\partial \Hscr_{sec,\,2+\frac{3}{2}}/\partial p_5=\omega_5\\
  \dot{\csi}_{1}=-\partial \Hscr_{sec,\,2+\frac{3}{2}} /\partial \eta_1=
  -\partial \big( \Hscr^{1-2}_{sec} + \Hscr^{1-3}_{sec} \big) /\partial \eta_1 \\
  \dot{\eta}_{1}=\partial \Hscr_{sec,\,2+\frac{3}{2}} /\partial \csi_1=
  \partial \big( \Hscr^{1-2}_{sec} + \Hscr^{1-3}_{sec} \big) /\partial \csi_1 \\
  \dot{P}_{1}=-\partial \Hscr_{sec,\,2+\frac{3}{2}} /\partial Q_1=
  -\partial \big( \Hscr^{1-2}_{sec} + \Hscr^{1-3}_{sec} \big) /\partial Q_1 \\
  \dot{Q}_{1}=\partial \Hscr_{sec,\,2+\frac{3}{2}}/\partial P_1=
  \partial \big( \Hscr^{1-2}_{sec} + \Hscr^{1-3}_{sec} \big) /\partial P_1
\end{cases}\, .
\end{equation}
In~\cite{masloc2023} the above secular quasi-periodic restricted
Hamiltonian model (with $2+3/2$ degrees of freedom) is introduced and
validated through the comparison with several numerical integrations
of the complete four-body problem, hosting planets $\b\,$, $\c\,$,
$\d\,$ of the \Ups{system}.

Recall also that, as already expressed in Remark~$3.1$
of~\cite{masloc2023}, the SQPR Hamiltonian $\Hscr_{sec,\, 2+ 3/2}$ is
invariant with respect to a particular class of rotations, admitting a
constant of motion that could be reduced, so to decrease by one the
number of degrees of freedom of the model (see
Section~\ref{sub:appl_KAM_2+3/2degree} below). In particular, it is
possible to verify the following invariance law:
\begin{equation}
\label{invariance}
\frac{\partial \Hscr_{sec,\, 2+3/2}}{\partial \gamma_1}+
\frac{\partial \Hscr_{sec,\, 2+3/2}}{\partial \theta_1}+
\frac{\partial \Hscr_{sec,\, 2+3/2}}{\partial q_3}+
\frac{\partial \Hscr_{sec, \,2+3/2}}{\partial q_4}+
\frac{\partial \Hscr_{sec, \,2+3/2}}{\partial q_5}=0,
\end{equation}
being $p_3 + p_4 + p_5 + \Gamma_1 +\Theta_1$
preserved.\footnote{According to Remark~$3.1$ of~\cite{masloc2023},
  condition~\eqref{invariance} can be stated also as follows:
  $$
  \vcenter{\openup1\jot\halign{
      &\hbox {$\displaystyle {#}$\hfil}\cr
  &-\frac{\partial \Hscr_{sec, \,2+3/2}}{\partial \csi_1}
  \frac{\partial \csi_1}{\partial \varpi_1}
  -\frac{\partial \Hscr_{sec,\, 2+3/2}}{\partial \eta_1}
  \frac{\partial \eta_1}{\partial \varpi_1}
  -\frac{\partial \Hscr_{sec, \,2+3/2}}{\partial P_1}
    \frac{\partial P_1}{\partial \Omega_1}
    -\frac{\partial \Hscr_{sec,\, 2+3/2}}{\partial Q_1}
    \frac{\partial Q_1}{\partial \Omega_1}
  \cr
  &+\frac{\partial \Hscr_{sec,\, 2+3/2}}{\partial q_3}
  +\frac{\partial \Hscr_{sec,\, 2+3/2}}{\partial q_4}
  +\frac{\partial \Hscr_{sec,\, 2+3/2}}{\partial q_5}=0.
  \cr
  }}
  $$
  }
This will be used in Section~\ref{sec:KAM}.

\section{Application of the normal form algorithm constructing elliptic tori to the SQPR model of the dynamics of \ups $\b$}
\label{sec:TE}

The
canonical change of variables $(\vet{p},\vet{q},\csi_1, \eta_1, P_1, Q_1)=\Ascr(\vet{p},\vet{q},\vet{I},\vet{\alpha})$ described by
\begin{equation}
\label{coord:I_alpha}
\begin{aligned}
&\csi_1=\sqrt{2 I_1}\cos(\alpha_1) \, , \qquad & &\eta_1=\sqrt{2 I_1}\sin(\alpha_1)\, ,\\
&P_1=\sqrt{2 I_2}\cos(\alpha_2)\, , \qquad & &Q_1=\sqrt{2 I_2}\sin(\alpha_2)\, ,
\end{aligned}
\end{equation} 
allows to rewrite the expansion of the SQPR
Hamiltonian~\eqref{Ham.b.new} as follows:
\begin{equation}
\label{Ham.T.E.0.the.prequel}
\begin{aligned}
  &\Hscr_{sec,\, 2+3/2}(\vet{p},\vet{q},\vet{I},\vet{\alpha})
  = \omega_3\,p_3 + \omega_4\,p_4 + \omega_5\,p_5
  \\
  & +\!\!\!\!\!\!\!\!\ \sum_{\substack{l_1+l_2=0\\(l_1\,,\,l_2)\in\mathbb{N}^2}}^{\mathcal{N}_L} \sum_{\substack{(k_3\,,\,k_4\,,\,k_5)\in\mathbb{Z}^3\\ |\vet{k}|\leq \mathcal{N}_SK}}
  \!\!\!\! \!\!\!\!\!\!\!\!\!\!\!\!\!\!\!\!\!\!\!\sum_{\substack{\qquad\qquad k_j=-l_j,-l_j+2,\ldots,l_j\\ j=1,\,2}}
  \!\!\!\!\!\!\!\!\!\!\!\!\!\!\!\!\!\!\!\!\! c_{\vet{l},\vet{k}} (\sqrt{I_1})^{l_1}(\sqrt{I_2})^{l_2}
  e^{i(k_1\alpha_1+k_2\alpha_2+k_3 q_3 + k_4 q_4 + k_5 q_5)}
 ,
\end{aligned}
\end{equation}
where $\vet{k}=(k_1,\ldots,k_5)\in\interi^5\,$ and the parameters
$\mathcal{N}_L$ and $\mathcal{N}_S$ define the truncation order of the
expansions in Taylor and Fourier series, respectively, in such a way
to represent on the computer just a finite number of terms that are
not too many to handle with. In our computations (without general
relativistic corrections) we fix $\mathcal{N}_L=6$ as maximal power
degree in square root of the actions and we include Fourier terms up
to a maximal trigonometric degree of $8$, putting $\mathcal{N}_S=4\,$,
$K=2\,$.  The r.h.s. of the above equation can be expressed in the
general and more compact form described by
\begin{equation}
\label{Ham_init_our}
\begin{aligned}
  \Hscr^{(0)}(\vet{p},\vet{q}, \vet{I},\vet{\alpha})=
  \Hscr_{sec,\, 2+3/2}(\vet{p},\vet{q},\vet{I},\vet{\alpha})
  &=
  \Escr^{(0)}+\vet{\omega}^{(0)}\cdot\vet{p}+\vet{\Omega}^{(0)}\cdot\vet{I}+
  \sum_{s= 0}^{\mathcal{N}_S}\sum_{l= 3}^{\mathcal{N}_L}
  f_{l}^{(0,s)}(\vet{q},\vet{I},\vet{\alpha})\\
  &\phantom{=}+\sum_{s= 1}^{\mathcal{N}_S}\sum_{l=0}^{2}
  f_{l}^{(0,s)}(\vet{q},\vet{I},\vet{\alpha}) \, ,
\end{aligned}
\end{equation}
whose structure is commented below. The algorithmic construction of
the normal form corresponding to an invariant elliptic torus starts
from the Hamiltonian $\Hscr_{sec,\, 2+3/2}$ rewritten in the same form
as $\Hscr^{(0)}$ in~\eqref{Ham_init_our} and its computational
procedure is fully detailed in Section~$4.1$ of~\cite{masloc2023},
where the approach explained in~\cite{locetal2022} is adapted to our
purposes (see also~\cite{carloc2021} for an application to the problem
of the FPU non-linear chains). In the present Section, this
constructive method is briefly summarized in the following way.

In order to perform the generic $r$-th step of the normalization
procedure, we first consider a Hamiltonian $\Hscr^{(r-1)}$ written as
follows:
\begin{equation}
\label{Ham.T.E.0}
\begin{aligned}
  \Hscr^{(r-1)}(\vet{p},\vet{q}, \vet{I},\vet{\alpha}) &=
  \Escr^{(r-1)}+\vet{\omega}^{(r-1)}\cdot\vet{p}+\vet{\Omega}^{(r-1)}\cdot\vet{I}+
  \sum_{s\geq 0}\sum_{l\geq 3} f_{l}^{(r-1,s)}(\vet{q},\vet{I},\vet{\alpha})
  \\
  &\phantom{=}
  +\sum_{s\geq r}\sum_{l=0}^{2} f_{l}^{(r-1,s)}(\vet{q},\vet{I},\vet{\alpha})\, ,
\end{aligned}
\end{equation}
where $\Escr^{(r-1)}$ is a constant term, with the physical dimension
of the energy, $(\vet{p},\vet{q})\in\reali^{n_1}\times\toro^{n_1}\,$,
$(\vet{I},\vet{\alpha})\in\reali^{n_2}_{\geq 0}\times \toro^{n_2}$ are
action-angle variables and
$(\vet{\omega}^{(r-1)},\vet{\Omega}^{(r-1)})\in\reali^{n_1}\times
\reali^{n_2}$ is an angular velocity vector (with $n_1=3$ and
$n_2=2\,$). The symbol $f_{l}^{(r-1,s)}$ is used to denote a function
of the variables $(\vet{q}, \vet{I},\vet{\alpha})$ which belongs to
the class $\mathfrak{P}_{l,sK}\,$, such that $l$ is the total degree
in the square root of the actions $\vet{I}$, $sK$ is the maximum of
the trigonometric degree, in the angles $(\vet{q},\vet{\alpha})\,$,
for a fixed positive integer $K$ (that is convenient to
put\footnote{Setting $K=2$ is quite natural for Hamiltonian systems
  close to stable equilibria, see, e.g.,~\cite{giolocsan2017}.} equal
to $2$ in the present context); finally, the upper index $r-1$ refers
to the number of normalization steps that have been already performed
during the execution of the constructive algorithm. Let us stress that
the Hamiltonian described in formula~\eqref{Ham_init_our} can be
rewritten in the form~\eqref{Ham.T.E.0} with $r=1$, by just splitting
the Fourier expansions in a suitable way.  Therefore, the algorithm
constructing the normal form for elliptic tori is applied by starting
the first normalizazion step from a Hamiltonian of the
type~\eqref{Ham.T.E.0}, with $r=1$, where the terms appearing in
the second row (namely, $\sum_{l=0}^{2}\sum_{s\geq 1}
f_{l}^{(0,s)}(\vet{q},\vet{I},\vet{\alpha})\,$) are considered as the
perturbation to remove. If the perturbing terms are small enough, they
can be eliminated through a sequence of canonical transformations,
leading the Hamiltonian to the following final form:
\begin{equation}
\label{Ham.T.E.goal}
\begin{aligned}
  \Hscr^{(\infty)}(\t{\vet{p}}, \t{\vet{q}}, \t{\vet{I}},\t{\vet{\alpha}})
  &= \Escr^{(\infty)}+\vet{\omega}^{(\infty)}\cdot\t{\vet{p}}+
  \vet{\Omega}^{(\infty)}\cdot\t{\vet{I}}+
  \sum_{s\geq 0}\sum_{l\geq 3}
  f_{l}^{(\infty,s)}(\t{\vet{p}},\t{\vet{q}},\t{\vet{I}},\t{\vet{\alpha}})\, ,
\end{aligned}
\end{equation}
with $f_{l}^{(\infty, s)}\in\mathfrak{P}_{l,sK}\,$. Therefore, for any
initial conditions of type $(\vet{0},\t{\vet{q}}_0, \vet{0},
\t{\vet{\alpha}})$ (where $\t{\vet{q}_0}\in\mathbb{T}^{n_1}$ and the
value of $\t{\vet{\alpha}}\in\mathbb{T}^{n_2}$ does not play any role,
since $(\t{\vet{I}},\t{\vet{\alpha}})=(\vet{0},\t{\vet{\alpha}})$
corresponds to $(\xi_1\,,\,\eta_1\,,\,P_1\,,\,Q_1)=(0,0,0,0)$
$\forall\ \t{\vet{\alpha}}$), the motion law
$(\t{\vet{p}}(t),\t{\vet{q}}(t),\t{\vet{I}}(t),\t{\vet{\alpha}}(t))=
(\vet{0},\t{\vet{q}}_0+\vet{\omega}^{(\infty)}t, \vet{0},
\t{\vet{\alpha}})$ is a solution of the Hamilton's equations related
to $\Hscr^{(\infty)}\,$. This quasi-periodic solution (having
$\vet{\omega}^{(\infty)}$ as constant angular velocity vector) lies on
the $n_1$-dimensional invariant torus such that the values of the
action coordinates are $\t{\vet{p}} = \vet{0},$
$\t{\vet{I}}=\vet{0}\,$.

The $r$-th normalization step consists of three substeps, each of them involving a canonical transformation which is expressed in terms of
the Lie series having $\chi_{0}^{(r)}$, $\chi_{1}^{(r)}$,
$\chi_{2}^{(r)}$ as generating function, respectively. Therefore, the
new Hamiltonian that is introduced at the end of the $r$-th
normalization step is defined as follows:
\begin{align}
\label{Ham.T.E.r.goal}
\Hscr^{(r)}=\exp\left(L_{\chi_2^{(r)}}\right)
\exp\left(L_{\chi_1^{(r)}}\right)\exp\left(L_{\chi_0^{(r)}}\right)
\Hscr^{(r-1)}\, ,
\end{align}
where $\exp \left( L_{\chi} \right)\cdot = \sum_{j \geq 0 }
(L_{\chi}^{j} \cdot)/j!$ is the Lie series operator, $ L_{\chi}
\cdot=\poisson{\cdot}{\chi}$ is the Lie derivative with respect to the
dynamical function $\chi\,$, and $\poisson{\cdot}{\cdot}$ represents
the Poisson bracket. The generating functions $\chi_0^{(r)}$,
$\chi_1^{(r)}$, $\chi_2^{(r)}$ are determined so as to remove the
angular dependence in the perturbing terms that are at most quadratic
in the square root of the actions $\vet{I}$ and of trigonometric
degree $rK$ in the angles $(\vet{q},\vet{\alpha})\,$, i.e., they
belong to the classes of functions $\mathfrak{P}_{0,rK}\,$,
$\mathfrak{P}_{1,rK}\,$ and $\mathfrak{P}_{2,rK}\,$, respectively.

Therefore, starting from~\eqref{Ham_init_our}, we perform
$\mathcal{N}_S$ normalization steps, bringing the Hamiltonian
$\Hscr^{(0)}$ in the following truncated normal form:
\begin{align}
\label{Ham.T.E.r.final.S}
\begin{aligned}
  \Hscr^{(\mathcal{N}_S)}(\vet{p},\vet{q}, \vet{I},\vet{\alpha}) &=
  \Escr^{(\mathcal{N}_S)}+\vet{\omega}^{(\mathcal{N}_S)}\cdot\vet{p} +
  \vet{\Omega}^{(\mathcal{N}_S)}\cdot\vet{I}+
  \sum_{s= 0}^{\mathcal{N}_S}\sum_{l= 3}^{\mathcal{N}_L}
  f_{l}^{(\mathcal{N}_S,s)}(\vet{q},\vet{I},\vet{\alpha})\, ,
\end{aligned}
\end{align}
where $f_{l}^{(\mathcal{N}_S,s)}\in\mathfrak{P}_{l,sK}$
$\forall\ l=3,\,\ldots\,,\,\mathcal{N}_L,\>s=0\,,\,\ldots\,,\,\mathcal{N}_S$
and the angular velocity vector related to the angles $\vet{q}$ is
such that
$\vet{\omega}^{(\mathcal{N}_S)}=\vet{\omega}^{(0)}=(\omega_3,
\omega_4, \omega_5)$, whose components are given
in~\eqref{freq.fond.CD}. This is due to the fact that the angular
velocity vector $\vet{\omega}$ does not change during the above
normalization procedure, since all the Hamiltonian terms, but
$\vet{\omega}\cdot\vet{p}\,$, do not depend on the dummy variables
$(p_3, p_4, p_5)\,$; thus, the components of $\vet{\omega}=(\omega_3,
\omega_4, \omega_5)\in\reali^3$ are written in
formula~\eqref{freq.fond.CD}, because they are still equal to the
values corresponding to the fundamental periods of the secular
dynamics of the outer exoplanets. Moreover, by the so called ``Exchange
Theorem'' (see~\cite{gro1967} and, e.g.,~\cite{gio2003}),
$\Hscr^{(\mathcal{N}_S)}(\vet{p},\vet{q},\vet{I},\vet{\alpha})=
\Hscr^{(0)}(\C^{(\mathcal{N}_S)}(\vet{p},\vet{q},\vet{I},\vet{\alpha}))$,
where
\begin{equation}
\label{cambio.coord.TE}
\begin{aligned}
\C^{(\mathcal{N}_S)}(\vet{p},\vet{q},\vet{I},\vet{\alpha})=
& \exp\left(L_{\chi_2^{(\mathcal{N}_S)}}\right)
  \exp\left(L_{\chi_1^{(\mathcal{N}_S)}}\right)
  \exp\left(L_{\chi_0^{(\mathcal{N}_S)}}\right)
  \ldots\\
  &\qquad\qquad\qquad\qquad\qquad\ldots\exp\left(L_{\chi_2^{(1)}}\right)
  \exp\left(L_{\chi_1^{(1)}}\right)\exp\left(L_{\chi_0^{(1)}}\right)
 (\vet{p},\vet{q},\vet{I},\vet{\alpha})\, .
\end{aligned}
\end{equation}

\section{KAM stability of $2+3/2$~DOF secular model of the innermost exoplanet orbiting in the $\upsilon$-Andromed{\ae} system}
\label{sec:KAM}

 In order to achieve our goal, it is convenient to adapt the
 Kolmogorov algorithm, in such a way to not keep fixed the angular
 velocity vector; such a slightly modified version of this classical
 normalization algorithm can be used in junction with a Newton-like
 method. As the main result of the present Chapter, we will show that
 the computational procedure explained in the following can be
 performed preliminarly in order to locate an approximation of the
 preselected orbit, which is a starting point good enough to
 successfully complete the CAP of existence of the wanted KAM torus.

\subsection{Reduction of the angular momentum}
\label{sub:appl_KAM_2+3/2degree}

After having performed $\mathcal{N}_S$ steps of the algorithm
constructing the normal form for an elliptic torus (according to the
prescriptions given in Subsection $4.1$ of~\cite{masloc2023} and
briefly recalled in the previous Section), the SQPR model of the
orbital dynamics of \ups$\b$ (taking into account, or not, the general
relativity corrections) is described by the Hamiltonian written
in~\eqref{Ham.T.E.r.final.S}, i.e.:
$$
\Hscr^{(\mathcal{N}_S)}(\vet{p},\vet{q}, \vet{I},\vet{\alpha})
=\Escr^{(\mathcal{N}_S)}+\vet{\omega}\cdot\vet{p}
+\vet{\Omega}^{(\mathcal{N}_S)}\cdot\vet{I}
+\sum_{s=0}^{\mathcal{N}_S}\sum_{l= 3}^{\mathcal{N}_L}
f_{l}^{(\mathcal{N}_S,s)}(\vet{q},\vet{I},\vet{\alpha})\, ,
$$
where the Taylor--Fourier expansion above is
truncated\footnote{\label{foot_param} Following~\cite{masloc2023}, we
  fix the parameters ruling the truncation of the series expansion so
  that $\mathcal{N}_L=6\,$, $\mathcal{N}_S=4$ and $K=2\,$ if we are
  not considering the GR correction on the SQPR model,
  $\mathcal{N}_L=6\,$, $\mathcal{N}_S=5$ and $K=2\,$ if we are dealing
  with the SQPR model with GR corrections.} up to order
$\mathcal{N}_L$ with respect to the square root of the actions
$\vet{I}$ and to the trigonometric degree $\mathcal{N}_S K$ in the
angles $(\vet{q},\vet{\alpha})$, while
$\Escr^{(\mathcal{N}_S)}\in\reali$ and the summands
$f_l^{(0,s)}\in\mathfrak{P}_{l, sK}\,$.

As already remarked at the end of Section~\ref{sec:SQPR}, the
Hamiltonian $\Hscr_{sec,\, 2+3/2}$ is invariant with respect to a
particular class of rotations; thus, since the Lie series introduced
in Section~\ref{sec:TE} preserve such invariance law, the same holds
for $\Hscr^{(\mathcal{N}_S)}$. Therefore, it is convenient to
reduce\footnote{In the previous Sections and in~\cite{masloc2023}, we
  have decided to not perform such a reduction, in order to make the
  role of the angular (canonical) variables more transparent, in such
  a way to clarify their meaning with respect to the positions of the
  exoplanets.} the number of DOF as we are going to explain. We
consider the canonical transformation
$(\vet{I},\vet{p},\vet{\alpha},\vet{q})=\F(P_1,P_2,P_3,P_4,P_5,Q_1,Q_2,Q_3,Q_4,Q_5)$,
expressed by the following generating function (in mixed coordinates):
$$
\vcenter{\openup1\jot\halign{
 \hbox {\hfil $\displaystyle {#}$}
 &\hbox {$\displaystyle {#}$\hfil}\cr
 &S(I_1,I_2,p_3,p_4,p_5,Q_1,Q_2,Q_3,Q_4,Q_5)=
 \cr
 &\ \qquad\qquad
 \big(I_1-I_{1_{\ast}}\big) Q_1+ \big(I_2-I_{2_{\ast}}\big) Q_2
 + p_3 Q_3 +p_4 Q_4 + (I_1 + I_2 - I_{1_{\ast}}-I_{2_{\ast}}+ p_3 + p_4 + p_5) Q_5\, ,
 \cr
}}
$$
where the translation vector
$\big(I_{1_{\ast}}\,,\,I_{2_{\ast}}\big)$ includes two constant
parameters that will be determined as explained in the next Sections.
The corresponding canonical change of variables is explicitely given
by
\begin{equation}
\label{change_loose_1DOF}
\begin{aligned}
&I_1=P_1+I_{1_{\ast}}\,, \qquad & & \alpha_1=Q_1+Q_5,\\
&I_2=P_2+I_{2_{\ast}}\,, \qquad & & \alpha_2=Q_2+Q_5, \\
&p_3=P_3, \qquad & & q_3=Q_3+Q_5, \\
&p_4=P_4, \qquad & & q_4=Q_4+Q_5, \\
&p_5=P_5-P_1-P_2-P_3-P_4, \qquad & & q_5=Q_5\, .
\end{aligned}
\end{equation}
After having expressed the Hamiltonian $\Hscr^{(\mathcal{N}_S)}$ as a
function of the new canonical coordinates $(P_1$, $P_2$, $P_3$, $P_4$,
$P_5$, $Q_1$, $Q_2$, $Q_3$, $Q_4$, $Q_5)$, then one can easily check
that\footnote{It is sufficient to apply the chain rule to
  expression~\eqref{invariance}. Moreover, recalling the changes of
  variables~\eqref{Poinc.var.U} and ~\eqref{coord:I_alpha} (leading to
  $I_1=\Gamma_1$, $I_2=\Theta_1$), it is easy to see that the
  constancy of $\Gamma_1+\Theta_1+p_3+p_4+p_5$ is equivalent to the
  one of $P_5\,$. }
$$
\frac{\partial \Hscr^{(\mathcal{N}_S)}}{\partial Q_5}=0\, ;
$$ in words, this is equivalent to say that $Q_5$ is a cyclic
angle \footnote{\label{soluz_Q5} Recall that $Q_5(t)=Q_5(0)+\omega_5
  \,t $, where $\omega_5$ is described in formula~\eqref{freq.fond.CD}.}
and, therefore, its conjugate momentum
$P_5=I_1-I_{1_{\ast}}+I_2-I_{2_{\ast}}+p_3+p_4+p_5$ is a constant of
motion. It is worth to add here some comments, in order to clarify the
role of the pair $(P_5\,,\,Q_5)$. Apart the modifications introduced
by all the near-to-identity canonical transformations\footnote{It is
  not difficult to verify that the all Lie series introduced in
  Section~\ref{sec:TE} preserve the invariance with respect to
  $Q_5\,$.} which define the normalization procedure described in
Section~\ref{sec:TE}, $q_3$ and $q_4$ correspond to the longitudes of
the pericenters of \ups$\c$ and \ups$\d$, respectively, while $q_5$
refers to the longitude of the nodes of \ups$\c$ and \ups$\d$ (that
are opposite each other, in the Laplace frame determined by taking
into account just these two exoplanets). This identification is due to
the way we have determined $(q_3,q_4,q_5)$ by decomposing some
specific signals of the secular dynamics of the outer exoplanets (this
is made by using the Frequency Analysis as it is sketched in
Section~\ref{sec:SQPR}). Moreover, $q_1=\alpha_1$ and $q_2=\alpha_2$
correspond to the longitude of the pericenter and the longitude of the
node of \ups$\b$, respectively. Therefore, it is not difficult to see
that the dynamics of the model we are studying does depend just on the
pericenters arguments of the three exoplanets and on the difference
between the longitude of the nodes of \ups$\b$ and \ups$\c$, i.e.,
$\Omega_1-\Omega_2=\Omega_1-\Omega_3-\pi$. Since the Hamiltonian is
invariant with respect to any rotation of the same angle that is
applied to all the longitudes of the nodes, then the total angular
momentum is preserved. Thus, $P_5$ is constant because it describes
the total angular momentum.

We focus our analysis on the 2+2/2~DOF Hamiltonian 
\begin{align*}
  & \Hscr^{(\mathcal{N}_S)}(P_1+I_{1_{\ast}},P_2+I_{2_{\ast}},P_3,P_4,
  P_5-P_1-P_2-P_3-P_4,Q_1+Q_5,Q_2+Q_5,Q_3+Q_5,Q_4+Q_5,Q_5)\,,
\end{align*}
where we have stressed the parametric role of the constant values of
$I_{{\ast}}=\big(I_{1_{\ast}}\,,\,I_{2_{\ast}}\big)$ and the angular
momentum $P_5$ is replaced by its constant value, which can be fixed
according to the initial conditions, i.e.,
$P_5=P_5(0)=I_1(0)-I_{1_{\ast}}+I_2(0)-I_{2_{\ast}}+p_3(0)+p_4(0)+p_5(0)$.
The Taylor expansion around $P_1=0$, $P_2=0$ of the previous
expression of $\Hscr^{(\mathcal{N}_S)}$ can be written as follows:
\begin{equation}
\label{Ham.Kam.moreDOF.iniz}
\Hscr_K^{(0)}(\vet{P},\vet{Q};\,\vet{I}_\ast) =
\Escr^{(0)}(\vet{I}_\ast)+\big(\vet{\omega}^{(0)}(\vet{I}_\ast)\big)\cdot\vet{P}
+\sum_{s=0}^{\t{\mathcal{N}}_S}\sum_{l= 2}^{\t{\mathcal{N}}_L}
f_{l}^{(0,s)}(\vet{P},\vet{Q};\vet{I}_\ast)
+\sum_{s=1}^{\t{\mathcal{N}}_S}\sum_{l= 0}^{1}
f_{l}^{(0,s)}(\vet{P},\vet{Q};\vet{I}_\ast)\, ,
\end{equation}
where $(\vet{P},\vet{Q}):= (P_1,P_2,P_3,P_4,Q_1,Q_2,Q_3,Q_4)\,$, and
$\vet{\omega}^{(0)}(\vet{I}_\ast)\in\reali^4$ is defined so that
\begin{equation}
\label{new_freq_lose}
\begin{aligned}
  &\omega_1^{(0)}(\vet{I}_\ast)=
  \frac{\partial\avg{\Hscr_K^{(0)}}_{\vet{Q}}}{\partial P_1}
  \Bigg|_{\substack{P_1=0\\ P_2=0}}\,, &
  &\omega_2^{(0)}(\vet{I}_\ast)=
  \frac{\partial\avg{\Hscr_K^{(0)}}_{\vet{Q}}}{\partial P_2}
  \Bigg|_{\substack{P_1=0\\ P_2=0}}\,, &
  &\omega_3^{(0)}=\omega_3-\omega_5\,, &
  &\omega_4^{(0)}=\omega_4-\omega_5\,.
\end{aligned}
\end{equation}
The parameter $\t{\mathcal{N}}_L$ denotes the order
of truncation with respect to the actions $\vet{P}$ and
$\t{\mathcal{N}}_S\t{K}$ is the maximal trigonometric degree in the
angles $\vet{Q}\,$.
Moreover,
$\Escr^{(0)}=\avg{\Hscr_K^{(0)}}_{\vet{Q}}\Big|_{\vet{P}=\vet{0}}\in\reali$ (i.e. $\Escr^{(0)}\in\mathcal{P}_{0,0}$)
and the summands can be rearranged so that
$f_l^{(0,s)}\in\mathcal{P}_{l,s\t{K}}\,$. For a fixed
positive integer $\t{K}$ and $\forall\ l\ge 0,\,s\ge 0$, the class of
functions $\mathcal{P}_{l,s\t{K}}$ is defined in such a way that
\begin{align}
  \label{def.pol.P}
\mathcal{P}_{l,s\t{K}}=\Bigg\lbrace f:
\mathbb{R}^n\times\mathbb{T}^n \rightarrow \mathbb{R}\, :\, f(\vet{P},\vet{Q})
=\sum_{\substack{\vet{j}\in\mathbb{N}^n\\ |\vet{j}|=l}}\sum_{\substack{\vet{k}\in\mathbb{Z}^n
    \\ |\vet{k}|\leq s\t{K}}}
c_{\vet{j},\vet{k}}\,\vet{P}^{\vet{j}} e^{i\,\vet{k}\cdot\vet{Q}} \Bigg\rbrace\, ,
\end{align}
with $n$ denoting the number of degrees of freedom (i.e., $n=4$ in the
model we are studying).  Since every coefficient
$c_{\vet{j},\vet{k}}\in\complessi$, then the following relation holds
true: $c_{\vet{j},-\vet{k}} = \bar{c}_{\vet{j},\vet{k}}\,$. Let us
also recall that in the symbol $f_{l}^{(0,s)}$ the first upper index
denotes the normalization step.

\subsection{Algorithmic construction of the Kolmogorov normal form without fixing the angular velocity vector}
\label{subsec:KAM_no_freq_fixed}

For most of the present Subsection, we are going to follow rather
closely the approach described in Section~2.3
of~\cite{locetal2022}. However, the following explanation has to be
detailed enough in order to make the subsequent application of the
Newton-like method well defined.  We are going to describe an
algorithm that aims to bring the Hamiltonian
\begin{equation}
\label{Ham.general.KAM}
\begin{aligned}
\Kscr^{(0)}(\vet{P},\vet{Q}) =
\Escr^{(0)}+\vet{\omega}^{(0)}\cdot\vet{P}
+\sum_{s\geq 0}\sum_{l\geq 2}
f_{l}^{(0,s)}(\vet{P},\vet{Q})
+\sum_{s\geq 1}\sum_{l= 0}^{1}
f_{l}^{(0,s)}(\vet{P},\vet{Q})
\end{aligned}
\end{equation}
with $f_{l}^{(0,s)}\in \mathcal{P}_{l, s\t{K}}$, in \textit{Kolmogorov
  normal form}; this means that we want to remove the last series
appearing in the expansion~\eqref{Ham.general.KAM}, i.e., $\sum_{s\geq
  1}\sum_{l= 0}^{1} f_{l}^{(0,s)}(\vet{P},\vet{Q})$ which has to be
considered smaller than the rest of the Hamiltonian. In fact, because
of the Fourier decay of $f_l^{(0,s)}\in\mathcal{P}_{l,s\t{K}}$ (whose
expansion is generically described in~\eqref{def.pol.P}), a suitable
choice of the positive integer parameter $\t{K}$ and an eventual
reordering of the monomials allow to write the
expansion~\eqref{Ham.general.KAM} (and~\eqref{Ham.Kam.moreDOF.iniz})
in such a way that $f_{l}^{(0,s)}=\Oscr(\varepsilon^s)\,$. Thus, the
goal is to lead the Hamiltonian to the following Kolmogorov normal
form:
\begin{align}
\label{Ham.Kam.goal}
\Kscr^{(\infty)}(\vet{P},\vet{Q})=
\Escr^{(\infty)}+\vet{\omega}^{(\infty)}\cdot\vet{P}+\Oscr(||\vet{P}||^2),
\end{align}
admitting, as solution, the quasi-periodic motion (having
$\vet{\omega}^{(\infty)}$ as angular velocity vector) on the invariant
torus corresponding to $\vet{P}=\vet{0}\,$.

In order to check how the structure of the classes of functions is
preserved by the normalization algorithm, the following statement
plays an essential role.
\begin{lemma}
\label{lemma:pol_KAM}
Let us consider two generic functions $g\in\mathcal{P}_{l,s\t{K}}$ and
$h\in\mathcal{P}_{m,r\t{K}}\,$, where $\t{K}$ is a fixed positive
integer number. Then
$$
\poisson{g}{h}=L_{h}\,g\,\in\,\mathcal{P}_{l+m-1,\,(r+s)\t{K}}
\qquad\forall\,l,\,m,\,s\,\in\,\mathbb{N}\, ,
$$ where we trivially extend the definition~\eqref{def.pol.P} in such a
way that $\mathcal{P}_{-1,\,s\t{K}}=\{0\}$ $\forall\ s\in\mathbb{N}$.
\end{lemma}

We describe the generic $r$-th normalization step, starting from the
Hamiltonian
\begin{equation}
\begin{split}
\label{Ham.Kam.moreDOF.r-1}
\Kscr^{(r-1)}(\vet{P},\vet{Q}) =
\Escr^{(r-1)}+\vet{\omega}^{(r-1)}\cdot\vet{P}+
\sum_{s\geq 0}\sum_{l\geq 2} f_{l}^{(r-1,s)}(\vet{P},\vet{Q})+
\sum_{s\geq r}\sum_{l= 0}^{1} f_{l}^{(r-1,s)}(\vet{P},\vet{Q})\,,
\end{split}
\end{equation}
where $(\vet{P},\vet{Q})$ are action-angle variables,
$\vet{\omega}^{(r-1)}\in\reali^n$, $\Escr^{(r-1)}\in\reali$ is an
energy value and $f_l^{(r-1,s)}\in\mathcal{P}_{l,s\t{K}}\,$. The first
upper index (i.e., $r-1$) denotes the number of normalization steps
that has been already performed; therefore, the
expansion~\eqref{Ham.general.KAM} of~$\Kscr^{(0)}$ is coherent with
the one of $\Kscr^{(r-1)}$ in~\eqref{Ham.Kam.moreDOF.r-1} when $r=1$.

In order
to bring the Hamiltonian in \textit{Kolmogorov normal form}, a
sequence of canonical transformations is performed with the aim to
remove the (small) perturbing terms, that are represented by the last
series appearing in the expansion~\eqref{Ham.Kam.moreDOF.r-1}.  Thus,
the $r$-th normalization step consists of two substeps, each of them
involving two canonical transformations, that are defined by Lie
series. Their generating functions are $\chi_1^{(r)}(\vet{Q})$ and
$\chi_2^{(r)}(\vet{P},\vet{Q})$, respectively; thus, the new
Hamiltonian at the end of the $r$-th normalization step is defined as
\begin{equation}
\label{H.r.H.r-1_KAM_more}
\Kscr^{(r)}=
\exp\left(L_{\chi_2^{(r)}}\right)\exp\left(L_{\chi_1^{(r)}}\right)\Kscr^{(r-1)}\, .
\end{equation}
More precisely,
$\Kscr^{(r)}(\vet{P},\vet{Q})=\Kscr^{(0)}(\mathcal{T}^{(r)}(\vet{P},\vet{Q}))\,$,
where
\begin{align}
\label{cambio.coord.KAM}
\mathcal{T}^{(r)}(\vet{P},\vet{Q})=
\exp\left(L_{\chi_2^{(r)}}\right)\exp\left(L_{\chi_1^{(r)}}\right)\ldots
\exp\left(L_{\chi_2^{(1)}}\right)\exp\left(L_{\chi_1^{(1)}}\right)(\vet{P},\vet{Q})\,.
\end{align}

\subsubsection*{First substep (of the r-th normalization step)}

The first substep aims to remove the perturbing term
$f_0^{(r-1,r)}\,$; thus, the first generating function
$\chi_{1}^{(r)}(\vet{Q})$ is determined solving the following
homological equation:
\begin{equation}
  \label{newdef.chi1.KAM}
  \poisson{\vet{\omega}^{(r-1)}\cdot\vet{P}}{\chi_{1}^{(r)}}
  +f_{0}^{(r-1,r)}(\vet{Q})=\avg{f_{0}^{(r-1,r)}}_{\vet{Q}}\ .
\end{equation}
After having expanded the perturbing term as
$f_{0}^{(r-1,r)}(\vet{Q})=\sum_{|\vet{k}|\leq r\t{K}}c_{\vet{k}}^{(r-1)}
e^{i\vet{k}\cdot\vet{Q}}\,$, one easily gets
\begin{equation*}
  \chi_{1}^{(r)}(\vet{Q}) = \sum_{0<|\vet{k}|\leq r\t{K}}
  \frac{c_{\vet{k}}^{(r-1)}}{i\,\vet{k}\cdot\vet{\omega}^{(r-1)}}
  e^{i\vet{k}\cdot\vet{Q}}\, ,
\end{equation*}
which is well defined provided that the non-resonance condition
$\vet{k}\cdot\vet{\omega}^{(r-1)}\neq 0$ is satisfied
$\forall\,\,\vet{k}\in\interi^n\setminus\{\vet{0}\}$ such that
$0<|\vet{k}|\leq r\t{K}\,$.  At the end of this first
normalization substep, (by the abuse of notation that is usual in the
Lie formalism, i.e., the new canonical coordinates are denoted with
the same symbols as the old ones) the intermediate Hamiltonian can be
written as follows:
\begin{align}
\label{H.hat.KAM}
  \widehat{\Kscr}^{(r)}=\exp L_{\chi_1^{(r)}}\Kscr^{(r-1)}=
  \Escr^{(r)}+\vet{\omega}^{(r-1)}\cdot\vet{P}+
  \sum_{s\geq 0}\sum_{l\geq 2} \widehat{f}_{l}^{(r,s)}(\vet{P},\vet{Q})+
  \sum_{s\geq r}\sum_{l= 0}^{1} \widehat{f}_{l}^{(r,s)}(\vet{P},\vet{Q})\, .
\end{align}
From a practical point of view, in order to define
$\widehat{f}_{l}^{(r,s)}$, it can be more comfortable to refer to a
definition structured in such a way to mimic more closely what is
usually done in any programming language. Thus, it can be convenient
to first define the new summands as the old ones, so that
$\widehat{f}_{l}^{(r,s)}=f_l^{(r-1,s)}$ $\forall \,l\geq 0\,$, $s\geq
0\,$. Hence, each term generated by Lie derivatives with respect to
$\chi_1^{(r)}$ is added to the corresponding class of functions. Thus,
by abuse, we redefine\footnote{From a practical point of view, if we
  have to deal with finite sums (as, for instance, in
  formula~\eqref{Ham.Kam.moreDOF.r-1}), such that the index $s$ goes
  up to a fixed order called $\t{\mathcal{N}}_S$, then we have to
  require also that $1\leq j\leq \min\left( l ,\lfloor
  (\t{\mathcal{N}}_S-s)/r \rfloor\right)\,$.}  these new symbols so
that
\begin{align*}
  \widehat{f}_{l-j}^{(r,s+jr)}
  \hookleftarrow
  \frac{1}{j!} L_{\chi_{1}^{(r)}}^j f_{l}^{(r-1, s)}
  \qquad\forall\,l\geq 1 ,\,\,1\leq j\leq  l,\,\,s\geq 0\, 
\end{align*}
where with the notation $a \hookleftarrow b$ we mean that the quantity $a$ is redefined so as to be equal $a + b\,$. It is easy to see that, since $\chi_1^{(r)}$ depends on $\vet{Q}$ only, then its Lie derivative decreases by $1$
the degree in $\vet{P}$, while the trigonometrical degree in the angles $\vet{Q}$ is increased by $r\t{K}$, because of
Lemma~\ref{lemma:pol_KAM}. By applying repeatedly Lemma~\ref{lemma:pol_KAM} to the formul{\ae}
above, one can easily verify that for all the Hamiltonian terms
appearing in the expansion~\eqref{H.hat.KAM} it holds true that
$\widehat{f}_{l}^{(r,s)}\in\mathcal{P}_{l,s\t{K}}\,$; moreover, it is
also very easy to check (by induction) that
$\widehat{f}_{l}^{(r,s)}=\Oscr(\varepsilon^s)$.
Moreover, in view of~\eqref{newdef.chi1.KAM} we also
set $\widehat{f}_{0}^{(r,r)}=0$ and we update the energy so that
$\Escr^{(r)}=\Escr^{(r-1)}+\avg{f_{0}^{(r-1,r)}}_{\vet{Q}}\,$. 

\subsubsection*{Second substep (of the r-th normalization step)}

The second substep aims to remove the perturbing term
$\widehat{f}_{1}^{(r,r)}\,$; thus, the generating function
$\chi_{2}^{(r)}(\vet{P},\vet{Q})$ can be determined by solving the
following homological equation:
\begin{equation}
\label{newdef.chi2.KAM}
  \poisson{\vet{\omega}^{(r-1)}\cdot\vet{P}}{\chi_{2}^{(r)}}+
  \widehat{f}_{1}^{(r,r)}(\vet{P},\vet{Q})=
  \avg{\widehat{f}_{1}^{(r,r)}}_{\vet{Q}}\, .
\end{equation}
Since in the first substep the non-resonance condition has been
assumed to be true, we get
\begin{equation*}
  \chi_{2}^{(r)}(\vet{P},\vet{Q})=
  \sum_{|\vet{j}|=1}\!\!\!\!\!\!\!\!\!\sum_{\qquad 0<|\vet{k}|\leq r\t{K}}
  \frac{c_{\vet{j},\vet{k}}^{(r)}}{i\,\vet{k}\cdot\vet{\omega}^{(r-1)}}\,
  \vet{P}^{\vet{j}} e^{i\vet{k}\cdot\vet{Q}}\, ,
\end{equation*}
where
$\widehat{f}_{1}^{(r,r)}=\sum_{|\vet{j}|=1}\,\sum_{|\vet{k}|\leq r\t{K}}
c_{\vet{j},\vet{k}}^{(r)}\,\vet{P}^{\vet{j}} e^{i\vet{k}\cdot\vet{Q}}\,$.
Thus, the Hamiltonian at the end of the $r$-th normalization step can
be written (by the usual abuse of notation on the new variables,
renamed as the old ones) as follows:
\begin{equation}
\label{Ham.Kam.moreDOF.r-1.end}
\begin{aligned}
  \Kscr^{(r)} &= \exp L_{\chi_2^{(r)}}\widehat{\Kscr}^{(r)}
  = \exp L_{\chi_2^{(r)}}\exp L_{\chi_1^{(r)}}\Kscr^{(r-1)}\\
  &=\Escr^{(r)}+\vet{\omega}^{(r)}\cdot\vet{P}+
  \sum_{s\geq 0}\sum_{l\geq 2} f_{l}^{(r,s)}(\vet{P},\vet{Q})+
  \sum_{s\geq  r+1}\sum_{l= 0}^{1} f_{l}^{(r,s)}(\vet{P},\vet{Q})\, ,
\end{aligned}
\end{equation}
where, first, we introduce $f_{l}^{(r,s)}=\widehat{f}_{l}^{(r,s)}$
$\forall \,l\geq 0\,$, $s\geq 0\,$ and then, by abuse, we
redefine\footnote{From a practical point of view, if we have to deal
  with finite sums such that the index $s$ goes up to a fixed order
  called $\t{\mathcal{N}}_S$, then we have to require also that $1\leq
  j\leq \lfloor (\t{\mathcal{N}}_S-s)/r \rfloor$ and, in the case of
  $f_{1}^{(r,jr)} \hookleftarrow \frac{1}{j!}
  L^{j}_{\chi_2^{(r)}}\left(\vet{\omega}^{(r-1)}\cdot\vet{P}\,\right)$,
  $1\leq j\leq \lfloor \t{\mathcal{N}}_S/r \rfloor$.} these new
symbols so that
\begin{equation*}
  f_{1}^{(r,jr)} \hookleftarrow
  \frac{1}{j!} L^{j}_{\chi_2^{(r)}}\left(\vet{\omega}^{(r-1)}\cdot\vet{P}\right)
  \quad{\rm and}\quad
  f_{l}^{(\,r,s+jr)} \hookleftarrow
  \frac{1}{j!} L_{\chi_{2}^{(r)}}^j \widehat{f}_{l}^{(r, s)}
  \quad\forall\,l\geq 0 ,\, j\geq 1 ,\,s\geq 0\, ,
\end{equation*}
since the Lie derivative with respect to $\chi_2^{(r)}$ does not
change the degree in $\vet{P}$, while it increases the trigonometrical
degree in the angles $\vet{Q}$ by $r\t{K}$.  In view of the previous
homological equation~\eqref{newdef.chi2.KAM} we also set
$f_{1}^{(r,r)}=0$ and we redefine the angular velocity vector so that
$$
\vet{\omega}^{(r)}\cdot\vet{P}=
\vet{\omega}^{(r-1)}\cdot\vet{P} + \avg{\widehat{f}_{1}^{(r,r)}}_{\vet{Q}}\, .
$$
Finally, if we denote with $(\vet{P}^{(r)},\vet{Q}^{(r)})$ the so
called normalized coordinates after $r$ normalization steps, i.e.,
$\Kscr^{(r)}(\vet{P}^{(r)},\vet{Q}^{(r)})\,$, then they are related
to the original ones (that are referring to $\Kscr^{(0)}\,$,
i.e. $(\vet{P}^{(0)},\vet{Q}^{(0)})=(\vet{P},\vet{Q})\,$) by the
following equation:
\begin{align}
\label{new_coord_KAM_U}
(\vet{P}^{(0)},\vet{Q}^{(0)})=
\exp\left(L_{\chi_2^{(r)}}\right)\exp\left(L_{\chi_1^{(r)}}\right)\ldots\exp\left(L_{\chi_2^{(1)}}\right)\exp\left(L_{\chi_1^{(1)}}\right)(\vet{P},\vet{Q})
\bigg|_{\substack{\vet{P}=\vet{P}^{(r)}\\ \vet{Q}=\vet{Q}^{(r)}}}\, ;
\end{align}
in the following we will adopt the symbol $\mathcal{T}^{(r)}$ to
denote the canonical transformation above, i.e.,
$(\vet{P}^{(0)},\vet{Q}^{(0)})=
\mathcal{T}^{(r)}(\vet{P}^{(r)},\vet{Q}^{(r)})$, that can be fully
justified by using repeteadly the Exchange Theorem.

It is now convenient to explain how the Kolmogorov normalization
algorithm can be used in junction with a Newton-like method.  For the
sake of definiteness, as a first approximation of the translation
vector $\vet{I}_\ast$ we are looking for, let us consider the values
at the time $t=0$ of the actions
$\vet{I}_{\ast}^{(0)}=\big(I_{1_{\ast}}(0)\,,\,I_{2_{\ast}}(0)\big)$
in correspondence with the initial conditions\footnote{Starting from
  the initial values of the orbital parameters that have been
  preselected, we can compute the corresponding actions values
  $\big(I_{1}(0)\,,\,I_{2}(0)\big)$ after the normalization procedure
  which is briefly described in Section~\ref{sec:TE}, that has been
  designed in order to construct a suitable elliptic torus.}  of the
selected orbit. After having fixed a translation vector
$\vet{I}_{\ast}^{(\nNewt)}$ (where the upper index $\nNewt$ just
counts the number of times the Newton method is iterated) we apply the
algorithm for the construction of the Kolmogorov normal form, starting
from the Hamiltonian $\Kscr^{(0)}$, described in
formula~\eqref{Ham.general.KAM}. From a practical point of view, such
an algorithm can be explicitly performed up to a finite number
$\bar{r}\,$ of normalization step. Thus, this part of the
computational procedure provides us the expansion of
$\Kscr^{(\bar{r})}(\vet{P},\vet{Q};\,\vet{I}_{\ast}^{(\nNewt)})$,
which is of the same type with respect to that described
in~\eqref{Ham.Kam.moreDOF.r-1.end}, but we rewrite it in such a way to
emphasize its parametric dependence on the translation vector, i.e.,
\begin{equation}
\label{Ham.KAM.rbar}
\begin{aligned}
\Kscr^{(\bar{r})}(\vet{P},\vet{Q};\,\vet{I}_{\ast}^{(\nNewt)}) &=
 \Escr^{(\bar{r})}(\vet{I}_{\ast}^{(\nNewt)})
 +\big(\vet{\omega}^{(\bar{r})}(\vet{I}_{\ast}^{(\nNewt)})\big)\cdot\vet{P}
 +\sum_{s\geq 0}\sum_{l\geq 2}
 f_{l}^{(\bar{r},s)}(\vet{P},\vet{Q};\,\vet{I}_{\ast}^{(\nNewt)})
\\
&\phantom{=}  +\sum_{s\geq \bar{r}+1}\sum_{l=0}^{1}
 f_{l}^{(\bar{r},s)}(\vet{P},\vet{Q};\,\vet{I}_{\ast}^{(\nNewt)})\, .
\end{aligned}
\end{equation}
The Hamiltonian above refers to an approximation of the final
invariant torus whose angular velocity vector is
$\vet{\omega}^{(\bar{r})}(\vet{I}_{\ast}^{(\nNewt)})$. Of course, if
the $\nNewt$-th numerical approximation $\vet{I}_{\ast}^{(\nNewt)}$ is
close enough to the translation vector $\vet{I}_{\ast}$ we are looking
for, then also
$\vet{\omega}^{(\bar{r})}(\vet{I}_{\ast}^{(\nNewt)})$ will
be close to the angular velocity vector, namely
$\vet{\omega}^{(*)}$, we are targeting. In
order to find a better approximation of the translation vector
$\vet{I}_{\ast}$ (and, consequently, of the preselected quasi-periodic
orbit) we can proceed by applying the Newton method. Thus, the
approximations of the initial translation vector are iteratively
computed so that
$$
{\vet{I}}_{\ast}^{(\nNewt)}=
\vet{I}_{\ast}^{(\nNewt-1)} + \Delta{\vet{I}}_{\ast}^{(\nNewt-1)}\,,
\qquad \nNewt\geq 1
$$
where the correction $\Delta{\vet{I}}_{\ast}^{(\nNewt-1)}$ is given
by the following refinement formula:
$$
\Delta \vet{\omega}({\vet{I}}_{\ast}^{(\nNewt-1)}) +
\mathcal{J}({\vet{I}}_{\ast}^{(\nNewt-1)}) \Delta{\vet{I}}_{\ast}^{(\nNewt-1)}=0\, ,
$$
where $\Delta \vet{\omega}({\vet{I}}_{\ast}^{(\nNewt-1)})=
\big(\omega_1^{(\bar{r})}({\vet{I}}_{\ast}^{(\nNewt-1)})-\omega_{1}^{(*)}\,,\,
\omega_2^{(\bar{r})}({\vet{I}}_{\ast}^{(\nNewt-1)})-\omega_{2}^{(*)}\big)$
and the $2\times 2$ Jacobian matrix
$\mathcal{J}({\vet{I}}^{(\nNewt-1)}_{\ast})$ of the function
${\vet{I}}_{\ast}^{(\nNewt-1)}\mapsto
\big(\omega_1^{(\bar{r})}({\vet{I}}_{\ast}^{(\nNewt-1)})\,,\,
\omega_2^{(\bar{r})}({\vet{I}}_{\ast}^{(\nNewt-1)})\big)$ is evaluated
numerically by the finite difference method\footnote{In practice, let
  us refer with the symbols
  $\vet{\omega}^{(\bar{r})}({\vet{I}}_{\ast}^{(\nNewt-1)})$,
  $\vet{\omega}^{(\bar{r})}(\t{\vet{I}}_1^{(\nNewt-1)})$,
  $\vet{\omega}^{(\bar{r})}(\t{\vet{I}}_2^{(\nNewt-1)})$ to the
  angular velocity vectors as they are determined at the end (i.e.,
  after $\bar{r}$ normalitation steps) of the Kolmogorov normalization
  algorithm which start from the initial translation vectors
  ${\vet{I}}_{\ast}^{(\nNewt-1)}=(I_{1_{\ast}}^{(\nNewt-1)},I_{2_{\ast}}^{(\nNewt-1)})\,$,
  $\t{\vet{I}}_1^{(\nNewt-1)}=(I_{1_{\ast}}^{(\nNewt-1)}+h_1\,,\,I_{2_{\ast}}^{(\nNewt-1)})\,$,
  $\t{\vet{I}}_2^{(\nNewt-1)}=(I_{1_{\ast}}^{(\nNewt-1)}\,,\,I_{2_{\ast}}^{(\nNewt-1)}+h_2)\,$,
  respectively; then
  $$
  \mathcal{J}({\vet{I}}_{\ast}^{(\nNewt-1)})=
  \begin{pmatrix}
  \frac{\omega_1^{(\bar{r})}(\t{\vet{I}}_1^{(\nNewt-1)})-
    \omega_1^{(\bar{r})}({\vet{I}}_{\ast}^{(\nNewt-1)})}{h_1}
  & \frac{\omega_1^{(\bar{r})}(\t{\vet{I}}_2^{(\nNewt-1)})
    -\omega_1^{(\bar{r})}({\vet{I}}_{\ast}^{(\nNewt-1)})}{h_2}\\
  \frac{\omega_2^{(\bar{r})}(\t{\vet{I}}_1^{(\nNewt-1)})
    -\omega_2^{(\bar{r})}({\vet{I}}_{\ast}^{(\nNewt-1)})}{h_1}
  &\frac{\omega_2^{(\bar{r})}(\t{\vet{I}}_2^{(\nNewt-1)})
    -\omega_2^{(\bar{r})}({\vet{I}}_{\ast}^{(\nNewt-1)})}{h_2}
  \end{pmatrix}\, .
  $$ In all our applications we have set the small increments in such
  a way that $h_1=I_{1_{\ast}}^{(\nNewt-1)}/100$ and
  $h_2=I_{2_{\ast}}^{(\nNewt-1)}/100\,$.} since an explicit analytic
expression of such a complicated function is not available. As usual,
the Newton method will be iterated until the
discrepancy\footnote{Hereafter, the norm $\|\cdot\|_{\infty}$ is
  defined so that $\|\vet{v}\|_{\infty}=\max_j|v_j|$
  $\forall\ \vet{v}\in\reali^m$ for any
  $m\in\naturali\setminus\{0\}$.} $\|\Delta
\vet{\omega}({\vet{I}}_{\ast}^{(\nNewt)})\|_{\infty}$ is smaller
than a prefixed tolerance threshold.

\subsection{Applications of the Kolmogorov normalization algorithm to the SQPR Hamiltonian model with $2+2/2$ DOF}
\label{subsec:results_KAM_no_freq_fixed}
Here, we want to construct the Kolmogorov normal form, as it has been
explained in the previous Section, taking as starting Hamiltonian
$\Kscr^{(0)}(\vet{P},\vet{Q})=\Hscr_K^{(0)}(\vet{P},\vet{Q};\,\vet{I}_\ast)$
defined in formula~\eqref{Ham.Kam.moreDOF.iniz}. In its expansion the
parametric dependency on the translation vector
$\vet{I}_{{\ast}}=\big(I_{1_{\ast}}\,,\,I_{2_{\ast}}\big)$ is emphasized for
each Hamiltonian summand.\footnote{Of course, in order to simplify the
  notation, such a dependency on the translation vector
  $\vet{I}_{{\ast}}=\big(I_{1_{\ast}}\,,\,I_{2_{\ast}}\big)$ has been
  omitted in the general description of
  Section~\ref{subsec:KAM_no_freq_fixed}.} The initial canonical
transformation~\eqref{change_loose_1DOF} is fully determined when the
components of $\vet{I}_{{\ast}}$ are fixed. In our strategy, we aim to
choose the values of $I_{1_{\ast}}$ and $I_{2_{\ast}}$ in such a way
that, at the end of the algorithm {\it \`a la} Kolmogorov, the wanted
angular velocity vector $\vet{\omega}^{(*)}$ is approached by the one
that is introduced at the end of each normalization step, i.e.,
$\vet{\omega}^{(r)}(\vet{I}_\ast)$ with
$r=0,\,1,\,\ldots\,\bar{r}\,$. We can determine
the angular velocity vector of the selected orbit by performing a
numerical integration of the original $2+3/2$ DOF Hamiltonian
model,\footnote{The equations of motion of the secular quasi-periodic
  restricted model are written in formula~\eqref{campo.Ham.b}.}
described in formula~\eqref{Ham.b.new} and applying the frequency
analysis method to the discretized signals
$t\mapsto\sqrt{2\Lambda_1}\,\sqrt{1-\sqrt{1-\big(\e_1(t)\big)^2}}
\,e^{-i\varpi_1(t)}\,$, $t\mapsto
2\sqrt{\Lambda_1}\,\sqrt[4]{1-\big(\e_1(t)\big)^2}
\,\sin\big(\frac{\i_1(t)}{2}\big)\,e^{-i\Omega_1(t)}$
(see~\eqref{Poinc.var.U} to recall the definition of the Poincar\'e
canonical variables). Let us denote with $\t{\omega}_{1}$ and
$\t{\omega}_{2}$ the values of the fundamental angular velocities
corresponding to these two signals, respectively. Taking into account
the canonical transformation~\eqref{change_loose_1DOF}, then we can
finally provide the values of the components of the angular velocity
vector $\vet{\omega}^{(*)}$, i.e.,
\begin{equation}
\label{freq_punto}
\omega_1^{(*)}=\t{\omega}_1-\omega_5\,,
\quad
\omega_2^{(*)}=\t{\omega}_2-\omega_5\,,
\quad
\omega_3^{(*)}=\omega_3-\omega_5\,,
\quad
\omega_4^{(*)}=\omega_4-\omega_5\,,
\end{equation}
where the values of $(\omega_3,\omega_4,\omega_5)\in\reali^3$ are
related to the fundamental periods of the two outer exoplanets and are
given in equation~\eqref{freq.fond.CD}. We remark that since the
actions $P_3$ and $P_4$ play the role of dummy variables during the
Kolmogorov normalization algorithm, just the first two components of
the angular velocity vector are updated at the end of every $r$-th
step of such a computational procedure, i.e.,
$\omega_1^{(r)}(\vet{I}_\ast)$ and\footnote{Let us recall that, in
  view of formul{\ae}~\eqref{freq_punto} and~\eqref{new_freq_lose},
  $\omega_3^{(\ast)}=\omega_3^{(0)}$ and
  $\omega_4^{(\ast)}=\omega_4^{(0)}$.}
$\omega_2^{(r)}(\vet{I}_\ast)$.  We can now apply the Kolmogorov
normalization algorithm in junction with a Newton-like method, as
explained in previous Section~\ref{subsec:KAM_no_freq_fixed}, starting
from the Hamiltonian $\Hscr_{K}^{(0)}$, defined in
formula~\eqref{Ham.Kam.moreDOF.iniz}, and performing
$\bar{r}=\t{\mathcal{N}}_S\,$ normalization steps by using {\tt
  Mathematica} as an algebraic manipulator. Thus, we obtain the
following truncated Hamiltonian
\begin{align}
\label{Ham.KAM.fine.2+3/2DOF}
\vcenter{\openup1\jot\halign{
 \hbox {\hfil $\displaystyle {#}$}
 &\hbox {$\displaystyle {#}$\hfil}\cr
 \Hscr_K^{(\t{\mathcal{{N}}}_S)}(\vet{P},\vet{Q};\,\vet{I}_{\ast}^{(\nNewt)}) &=
 \Escr^{(\t{\mathcal{{N}}}_S)}(\vet{I}_{\ast}^{(\nNewt)})
 +\big(\vet{\omega}^{(\t{\mathcal{{N}}}_S)}(\vet{I}_{\ast}^{(\nNewt)})\big)\cdot\vet{P}
 +\sum_{s=0}^{\t{\mathcal{N}}_S}\sum_{l= 2}^{\t{\mathcal{N}}_L}
 f_{l}^{(\t{\mathcal{{N}}}_S,s)}(\vet{P},\vet{Q};\,\vet{I}_{\ast}^{(\nNewt)})
 \cr
 &= \Escr^{(\t{\mathcal{{N}}}_S)}(\vet{I}_{\ast}^{(\nNewt)})
 +\big(\vet{\omega}^{(\t{\mathcal{{N}}}_S)}(\vet{I}_{\ast}^{(\nNewt)})\big)\cdot\vet{P}
 +\Oscr\big(||\vet{P}||^2\big)\, ,
 \cr
}}
\end{align}
which is in Kolmogorov normal form. Indeed, it refers to an
(approximately) invariant torus, whose energy level is equal to
$\Escr^{(\t{\mathcal{{N}}}_S)}(\vet{I}_{\ast}^{(\nNewt)})$ and the
corresponding angular velocity vector is
$\vet{\omega}^{(\t{\mathcal{{N}}}_S)}(\vet{I}_{\ast}^{(\nNewt)})=
\big(\omega_1^{(\t{\mathcal{{N}}}_S)}(\vet{I}_{\ast}^{(\nNewt)})\,,\,
\omega_2^{(\t{\mathcal{{N}}}_S)}(\vet{I}_{\ast}^{(\nNewt)})\,,\,
\omega_3^{(*)}\,,\,\omega_4^{(*)}\big)$, where the last two components
are defined in formula~\eqref{freq_punto}. In order to find
$\vet{I}_{\ast}^{(\nNewt)}$, we iterate the Newton method until
$\|\Delta \vet{\omega}({\vet{I}}_{\ast}^{(\nNewt)})\|_{\infty}$ is
smaller than a prefixed tolerance threshold. Therefore, if such a
condition is reached, we will have constructed a Kolmogorov normal
form corresponding to an invariant torus approximating the preselected
quasi-periodic orbit in a so accurate way that $\big(
\omega_1^{(\t{\mathcal{{N}}}_S)}({\vet{I}}_{\ast}^{(\nNewt)})\,,\,
\omega_2^{(\t{\mathcal{{N}}}_S)}({\vet{I}}_{\ast}^{(\nNewt)})\big)
\simeq\big(\omega_1^{(*)}\,,\,\omega_2^{(*)}\big)$. In our application
the initial approximation provided by the initial conditions, i.e.
$\vet{I}_{\ast}^{(0)}=\big(I_{1_{\ast}}(0)\,,\,I_{2_{\ast}}(0)\big)$,
is good enough to successfully perform the Newton method that stops
regularly with a final discrepancy that gets smaller than the
tolerance threshold, which is fixed so to be equal to $10^{-10}$.

\section{Invariant tori in the SQPR Hamiltonian model with $2+2/2$ DOF}
\label{sec:CAP-1stResult}

\subsection{Results produced by the semi-analytic integration of the non-relativistic Hamiltonian model}
\label{subsub:risut_KAM}
In order to compare the numerical procedure with the semi-analytical
one, we start with the numerical integration of the SQPR Hamiltonian
model with $2+3/2$ DOF, whose corresponding equations of motion are
reported in formula~\eqref{campo.Ham.b}. As values of the initial
orbital parameters we choose $a_1(0)\,$, $\e_1(0)\,$, $M_1(0)$ and
$\omega_1(0)\,$ as reported in Table~\ref{tab:param.orb.b}.  For this
study we decide to consider the case with $(\i_1(0),
\Omega_1(0))=(17^{\circ},5^{\circ})$ because it approximately
corresponds to the center of the stability region for what concerns
the orbital dynamics of \ups$\b$, according to the numerical
explorations discussed in Section~$3$ of~\cite{masloc2023}.

\begin{table}[h]
\begin{minipage}{0.5\textwidth}
\begin{center}
\begin{tabular}{ll}
\toprule
 & $\upsilon$-And $\bf{b}$  \\
\midrule
$m \, [M_J]$ & $0.674$ \\
$a(0) \, [{\rm AU}]$ & $0.0594$  \\
$\e(0)$ & $0.011769$ \\
$\i(0)\, [^{\circ}]$ & $17.$ \\
$M(0)\, [^{\circ}]$ & $103.53$ \\
$\omega(0)\, [^{\circ}]$ & $51.14$ \\
$\Omega(0)\, [^{\circ}]$ & $5.$ \\
\bottomrule
\end{tabular}
\end{center}
\end{minipage}
\begin{minipage}{0.5\textwidth}
\caption{Values of the initial orbital parameters for \ups$\b$. The
  chosen value of the mass is the minimal one according
  to~\cite{mcaetal2010}. The values $a_1(0)\,$, $\e_1(0)\,$,
  $M_1(0)$ and $\omega_1(0)$ are reported from the stable prograde trial
  PRO2 of~\cite{deietal2015} (Table~$3\,$). The values of initial
  inclination and longitude of the node (denoted with $\i_1(0)$ and
  $\Omega_1(0)$, respectively) are taken from~\cite{masloc2023}; see
  the text for more details. The corresponding initial orbital
    parameters in the Laplace reference frame can be easily determined
    (see, e.g.,~\cite{mas2023}).}
\label{tab:param.orb.b}
\end{minipage}
\end{table}

Concerning the semi-analytical approach, we start from the $2+2/2$DOF
Hamiltonian, namely \eqref{Ham.Kam.moreDOF.iniz}. For the construction
of the Kolmogorov normal form (without fixing the angular velocity
vector) explained in Section~\ref{subsec:KAM_no_freq_fixed} and
applied in the previous
Section~\ref{subsec:results_KAM_no_freq_fixed}, we adopt
$\t{\mathcal{N}}_ L = 2\,$, $\t{K}= 2\,$, $\t{\mathcal{N}}_S = 6\,$ as
parameters ruling the truncations of the expansions. In particular,
$\nNewt=3$ iterations of the Newton method are enough to reach the
condition
$\|\vet{\omega}^{(\t{\mathcal{{N}}}_S)}({\vet{I}}_{\ast}^{(3)})
-\vet{\omega}^{(*)}\|_{\infty}<10^{-10}$, which allows us to
successfully conclude the search for the initial translation vector
that approximates well enough the one we are looking for, namely the
unknown ${\vet{I}}_{\ast}\,$. This means that we start from
$$
\Hscr_K^{(0)}(\vet{P},\vet{Q};\,\vet{I}_\ast^{(3)}) =
\Escr^{(0)}(\vet{I}^{({3})}_\ast)+
\big(\vet{\omega}^{(0)}(\vet{I}^{({3})}_\ast)\big)\cdot\vet{P}
+\sum_{s=0}^{\t{\mathcal{N}}_S}\sum_{l= 2}^{\t{\mathcal{N}}_L}
f_{l}^{(0,s)}(\vet{P},\vet{Q};\vet{I}_\ast^{({3})})
+\sum_{s=1}^{\t{\mathcal{N}}_S}\sum_{l= 0}^{1}
f_{l}^{(0,s)}(\vet{P},\vet{Q};\vet{I}_\ast^{({3})})\, ,
$$
with $\t{\mathcal{N}}_ L = 2\,$, $\t{K}= 2\,$, $\t{\mathcal{N}}_S =
6\,$ and, after $\t{\mathcal{N}}_S $ normalization steps, we arrive at
the Hamiltonian~\eqref{Ham.KAM.fine.2+3/2DOF}, i.e.,
\begin{align*}
\Hscr_K^{(\t{\mathcal{{N}}}_S)}(\vet{P},\vet{Q};\,\vet{I}_{\ast}^{({3})}) 
=
 \Escr^{(\t{\mathcal{{N}}}_S)}(\vet{I}_{\ast}^{({3})})
 +\big(\vet{\omega}^{(\t{\mathcal{{N}}}_S)}(\vet{I}_{\ast}^{({3})})\big)\cdot\vet{P}
 +\Oscr(||\vet{P}||^2)\,.
\end{align*}

\begin{figure}[!h]
\begin{minipage}{.5\textwidth}
\includegraphics[scale=0.36]{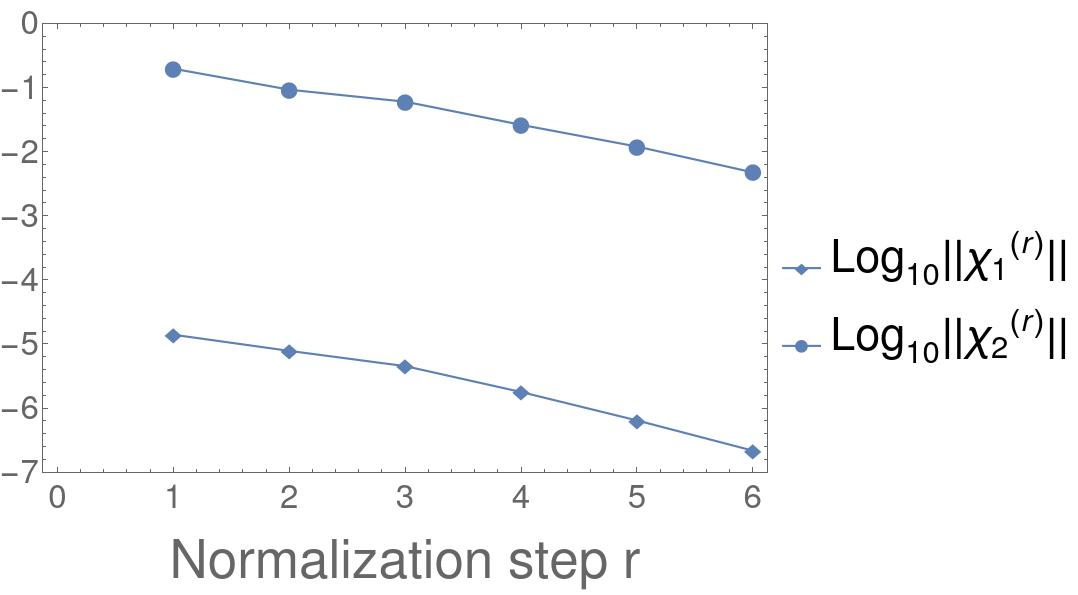}
\end{minipage}
\qquad\qquad\qquad\qquad
\begin{minipage}{.28\textwidth}
\caption{Convergence of the generating functions $\chi_1^{(r)}$ and
  $\chi_2^{(r)}$ defined by the normalization algorithm {\`a la}
  Kolmogorov without keeping fixed the angular velocity vector, when
  it is applied to the 2+2/2~DOF SQPR model {\it without} GR
  corrections in the case corresponding to $(\i_1 (0), \Omega_1 (0)) =
  (17^{\circ} , 5^{\circ} )$. The ${\rm Log}_{10}$ of their norms are
  reported as a function of the normalization step $r\,$.}
\label{fig.norm_decay_2+3/2}
\end{minipage}
\end{figure}

We focus on the last Kolmogorov normalization that is performed at the
end of the Newton method, i.e., the one corresponding to the initial
translation vector ${\vet{I}}_{\ast}^{({3})}$. In spite of the fact
that {\tt Mathematica} allows to deal just with a few normalization
steps in the case of a system with $2+2/2$ DOF, looking at
Figure~\ref{fig.norm_decay_2+3/2}, one can appreciate that the decay
of the norms of the generating functions $\chi_1^{(r)}$ and
$\chi_2^{(r)}$ is rather regular and sharp, where, for any function
$f(\vet{P},\vet{Q})\in\mathcal{P}_{l,s\t{K}}$ whose Taylor-Fourier
expansion is written in formula~\eqref{def.pol.P}, we define its norm
as
\begin{align*}
  \|f\|=
  \sum_{\substack{\vet{j}\in\mathbb{N}^n\\ |\vet{j}|=l}}
  \sum_{\substack{\vet{k}\in\mathbb{Z}^n
    \\ |\vet{k}|\leq s\t{K}}}
|c_{\vet{j},\vet{k}}|\, .
\end{align*}
This numerical evidence suggest that the normalization
algorithm should be convergent for $r\to\infty$.

\begin{figure}[!h]
\begin{minipage}{.45\textwidth}
\includegraphics[scale=0.28]{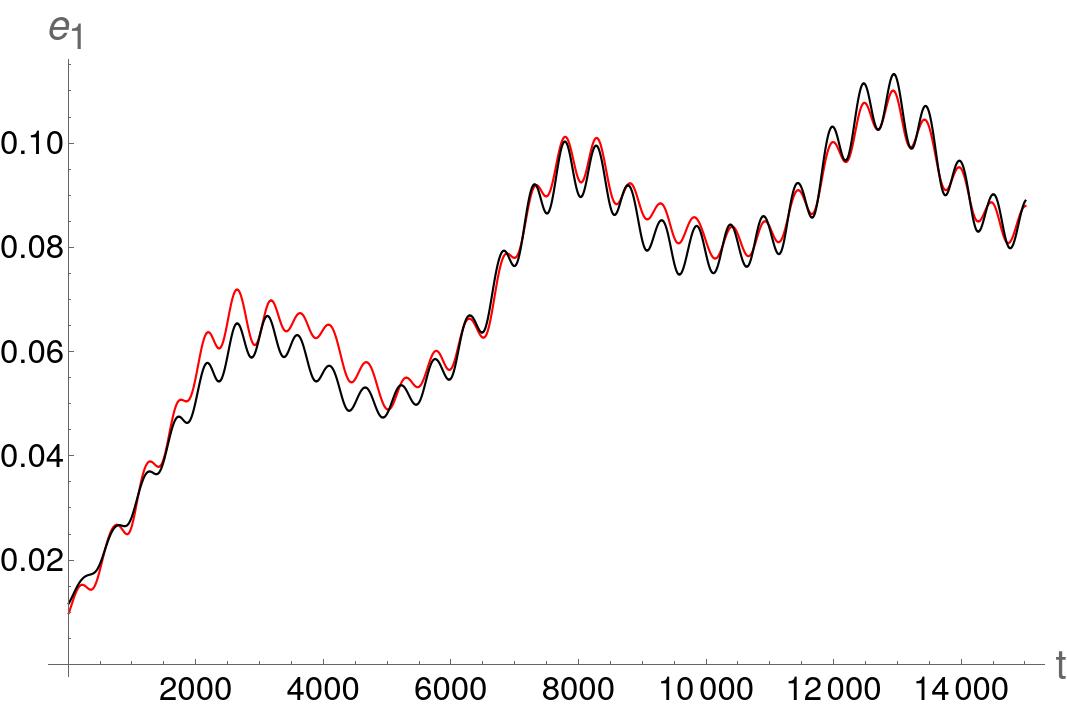}
\end{minipage}
\quad\quad
\begin{minipage}{.45\textwidth}
\includegraphics[scale=0.28]{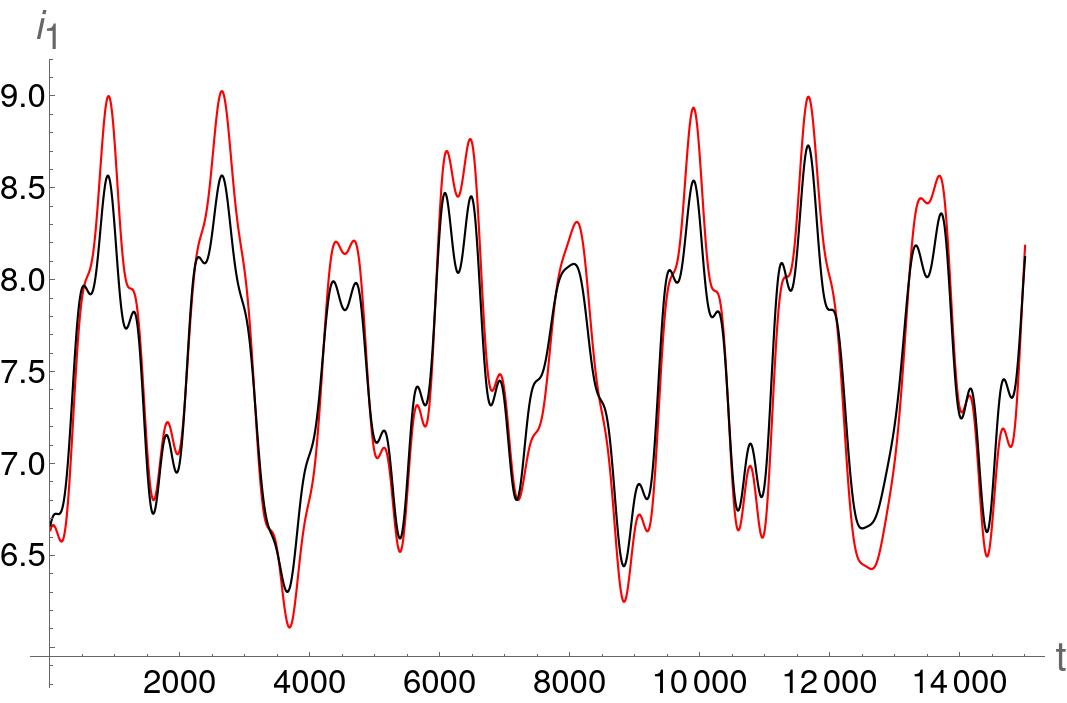}
\end{minipage}
\caption{Comparisons between the eccentricity $\e_1$ (on the left) and
  the inclination $\i_1$ (on the right) as obtained through the
  semi-analytical approach (in red) and the numerical one (in black).
  Both the integration methods consider the case with initial
  conditions referring to the data reported in
  Table~\ref{tab:param.orb.b} and they are applied to $2+3/2$~DOF
  model which {\it does not} take into account the effects due to GR.
  Units of measure for what concerns inclination $\i_1$ and time $t$
  are degree and year, respectively.}
\label{fig.comparison_e1_i1_2+3/2}
\end{figure}

Moreover, we can express the change of canonical coordinates allowing
us to recover the original variables of our problem, i.e., the ones
appearing as arguments of the Hamiltonian $\Hscr_{sec,\,2+3/2}\,$,
whose expansion is reported in~\eqref{Ham.T.E.0.the.prequel}. In fact,
they can be given as a function of the normalized canonical variables
that are listed in the final form of the
Hamiltonian~\eqref{Ham.KAM.fine.2+3/2DOF}. Let us recall that this
computational procedure ``just'' requires to compose all the canonical
transformations we have previously introduced. Therefore, by using
{\tt Mathematica} as an algebraic manipulator, it has been possible to
compute the expansions of such a composition of canonical
transformations described in formul{\ae}~\eqref{cambio.coord.TE},
\eqref{change_loose_1DOF} and~\eqref{cambio.coord.KAM}, in such a way
to to construct a semi-analytic (approximate) solution of the
equations of motion~\eqref{campo.Ham.b}.  This corresponds to the
invariant KAM torus related to the Kolmogorov normal
form~\eqref{Ham.KAM.fine.2+3/2DOF} which is travelled by
quasi-periodic orbits characterized by the angular velocity vector
$\vet{\omega}^{(*)}\,$, i.e., the motion law can be written as
$t\mapsto\big(\vet{p}(t),\vet{q}(t),\vet{I}(t),\vet{\alpha}(t)\big)=
\C^{(\mathcal{N}_s)}\big(\F\big(
\mathcal{T}^{(\t{\mathcal{N}}_s)}(\vet{0},\vet{Q}(0)+\vet{\omega}^{(*)}t),
  P_5(0), Q_5(0)+\omega_5\,t\big)\big)$.  Such a quasi-periodic
evolution is plotted (in red) in
Figure~\ref{fig.comparison_e1_i1_2+3/2}, where one can appreciate the
rather good agreement with the numerical integrations of the motion
(in black) for what concerns the behavior of both the eccentricity and
the inclination of \ups$\b$.

\subsection{Computer-Assisted Proofs of existence of KAM tori for the SQPR Hamiltonian models with $2+2/2$ DOF (without GR corrections)}
\label{sec:CAP}

Since we have been able to explicitly perform the normalization
procedure {\`a la} Kolmogorov just for very few steps of the algorithm
by using {\tt Mathematica}, the expectation of its convergence is not
supported in a convincing way by the results discussed in the previous
Subsection~\ref{subsub:risut_KAM}.  Therefore, we think it is
particularly interesting to adopt an approach based on a rigorous
Computer-Assisted Proof (hereafter, CAP) for the model introduced
above.  For such a purpose, it is convenient to run the code {\tt
  CAP4KAM{\_}nDOF} which can be downloaded from a publicly available
website${{\ref{footnote-CAP4KAM_nDOF}}\atop{\phantom{1}}}$; such a
software package is designed to prove the existence of KAM tori for
Hamiltonian systems with a number of DOF $n\ge 2$ (while the previous
similar version, namely {\tt CAP4KAM2D}\footnote{Available at the
  website \url{ https://doi.org/10.17632/jdx22ysh2s.1}}, is limited to
the case with 2~DOF). For what concerns the systems described in
Subsection~\ref{subsub:risut_KAM}, we apply the CAP to the new initial
Hamiltonian
\begin{equation}
\label{Ham.initial.CAP}
\begin{aligned}
  &H^{(0)}(\vet{P},\vet{Q})=
  \Hscr^{(\t{\mathcal{{N}}}_S-1)}(\vet{P},\vet{Q};\,\vet{I}_{\ast}^{({3})}) 
\\
&= \Escr^{(\t{\mathcal{{N}}}_S-1)}(\vet{I}_{\ast}^{({3})})
 +\big(\vet{\omega}^{(\t{\mathcal{{N}}}_S-1)}(\vet{I}_{\ast}^{({3})})\big)\cdot\vet{P}
 +\sum_{s=0}^{\t{\mathcal{N}}_s}\sum_{l= 2}^{\t{\mathcal{N}}_L}
 f_{l}^{(\t{\mathcal{{N}}}_S-1,\,s)}(\vet{P},\vet{Q};\,\vet{I}_{\ast}^{({3})})+
 \sum_{l= 0}^{1}
 f_{l}^{(\t{\mathcal{{N}}}_S-1,\,\t{\mathcal{N}}_S)}(\vet{P},\vet{Q};\,\vet{I}_{\ast}^{({3})})
 \,,
\end{aligned}
\end{equation}
with $\t{N}_L=2$, $\t{K}=2$ and $\t{N}_S=7\,$,\footnote{In this case
  we have put $\t{N}_S=7$ so that, as explained in the previous
  Section,
  $\vet{\omega}^{(6)}({\vet{I}}_{\ast}^{({3})})\simeq\vet{\omega}^{(*)}$,
  since $\|\big(
  \omega_1^{(6)}({\vet{I}}_{\ast}^{({3})})-\omega_{1}^{(*)}\,,\,
  \omega_2^{(6)}({\vet{I}}_{\ast}^{({3})})-\omega_{2}^{(*)}\big)\|_{\infty}<10^{-10}$. }
that corresponds to the Hamiltonian described in
formula~\eqref{Ham.KAM.rbar}, with $\bar{r}=\t{\mathcal{N}}_S-1=6$,
truncated up to degree $2$ in the actions and to
trigonometrical degree $14$ in the angles. This means that we are
studying the last Kolmogorov algorithm started at the end of the
Newton method, but we consider the Hamiltonian produced at the end of
the {\it next to last normalization step}. Performing also the last
step, of course, would completely remove all the perturbing terms
that are represented in our {\it truncated} expansions (as already
done in Section~\ref{subsub:risut_KAM}, in order to produce
Hamiltonian~\eqref{Ham.KAM.fine.2+3/2DOF}); in such a case an
application of a CAP to $\Hscr_{K}^{(\t{\mathcal{{N}}}_S)}$ would be
completely pointless.  Stopping the preliminary algebraic
manipulations that are performed by using {\tt Mathematica} at the
next to last step allows us to consider the main perturbing terms that
would make part also of an {\it infinite} series expansion of
$\Hscr^{(\t{\mathcal{{N}}}_S-1)}$; therefore, in our opinion this
Hamiltonian is a significant starting point and the subsequent
normalization procedure is quite challenging.

In the case of the $2+2/2$ DOF Hamiltonian model
$\Hscr^{(\t{\mathcal{{N}}}_S-1)}(\vet{P},\vet{Q};\,\vet{I}_{\ast}^{({3})})$
the CAP succeeds in rigorously proving the existence of a set (with
positive Lebesgue measure) of KAM tori whose corresponding angular
velocity vectors are Diophantine and in an extremely small
neighborhood of $\vet{\omega}^{(*)}\,$. In the code {\tt
  CAP4KAM{\_}nDOF}, that automatically performs the CAP of existence of
KAM tori, there are two internal parameters playing a fundamental
role, called $R_{\rm{I}}$ and $R_{\rm{II}}$. $R_{\rm{I}}$ refers to
the number of normalization steps for which the expansions of the
generating functions are explicitly computed; this means that the code
explicitly performs $R_{\rm{I}}$ normalization steps of a classical
formulation of the Kolmogorov normalization algorithm, computing
$H^{(R_{\rm{I}})}$. As a main difference with respect to the
computational procedure explained in
Subsection~\ref{subsec:KAM_no_freq_fixed}, the classical formulation
of the Kolmogorov normalization algorithm takes also into account, at
each step $r$, another generating function
$\vet{\xi}^{(r)}\cdot\vet{Q}\,$, in order to keep fixed the wanted
angular velocity vector of the quasi-periodic motion on the final
torus, namely, $\vet{\omega}^{(*)}\,$ (see the original
work~\cite{kol1954} written by Kolmogorov
and~\cite{BenGGS-1984},~\cite{gioloc1997} for a more modern
reformulation that is based on the Lie series method).  Instead,
$R_{\rm{II}}$ corresponds to the last step for which just the upper
bounds of the generating functions are estimated; more precisely, the
size of the perturbation is reduced just iterating the estimates of
the norms of the terms of order $r$ with $R_{\rm{I}} < r \leq
R_{\rm{II}}$. The impact of the choice of the parameters $R_{\rm{I}}$
and $R_{\rm{II}}$ on the performances of this kind of CAPs is widely
discussed in~\cite{celetal2000}.  In the former case the values of the
parameters that are internal to the CAP and that mostly affect the
computational complexity are fixed so that $R_{\rm{I}}=42$ and
$R_{\rm{II}}=36000$. The CPU-time needed to complete the CAP is about
$74.9$~days on a workstation equipped with processors of type
\texttt{Intel XEON-GOLD 5220} (2.2~GHz).

\begin{figure}[!h]
\begin{minipage}{.48\textwidth}
\includegraphics[scale=0.68]{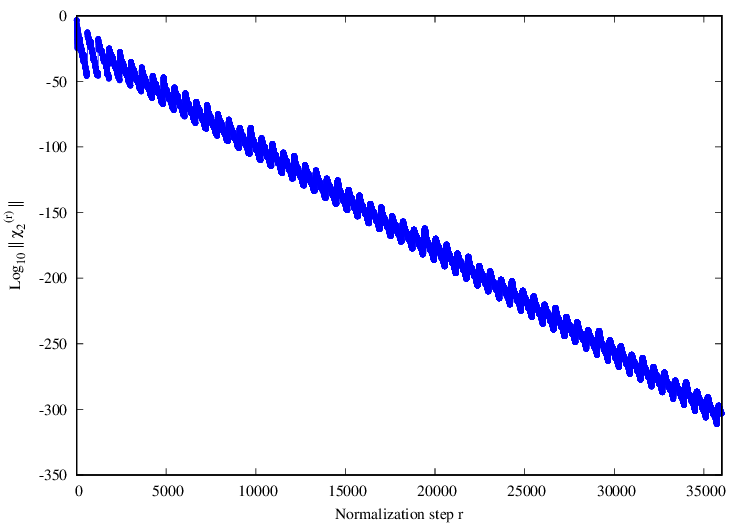}
\end{minipage}
\quad
\begin{minipage}{.48\textwidth}
\includegraphics[scale=0.68]{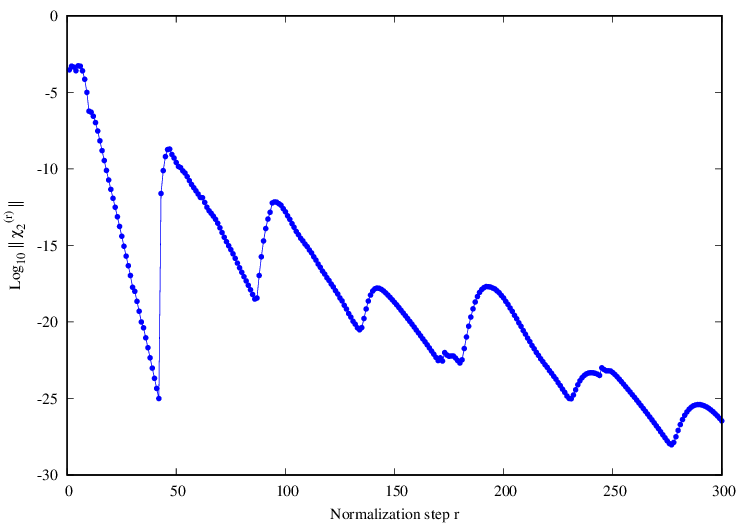}
\end{minipage}
\caption{Estimates of the norms (in semi-log scale) of the generating function $\chi_2^{(r)}$, as produced during the CAP, using $R_{\rm{I}}=42$ and $R_{\rm{II}}=36000$. On the right, the zoom on the first $300$ normalization steps.}
\label{fig.CAP}
\end{figure}

The plot (in semi-log scale) of the estimates of the norm of
$\chi_2^{(r)}$, reported in the left panel of Fig.~\ref{fig.CAP},
shows a regular decreasing behaviour. Moreover, looking at its zoom,
i.e., the plot on the right panel, it is evident that such a decrease
is steeper up to the step $R_{\rm{I}}$. Then there is a very
remarkable leap, due to the transition from the regime of the explicit
computation of the generating functions to the one of the pure
iteration of the estimates. For increasing values of the normalization
step $r$ it is evident the decreasing behaviour of the norm of
$\chi_2^{(r)}$, with some periodic small jumps.  By comparing the plot
in the left panel of Fig.~\ref{fig.CAP} with the one on the right, one
can easily realize that the behavior of the jumps is similar, but with
a different periodicity that is actually induced by another parameter
which is internal to the CAP, namely {\tt MAXMODKCALC}. It has been
fixed so as to be equal to~$600$; this means that the minimum of the
small divisors produced by the solution of the homological equations
(see formul{\ae}~\eqref{newdef.chi1.KAM} and~\eqref{newdef.chi2.KAM})
is explicitly computed for Fourier harmonics $\vet{k}$ such that
$|\vet{k}|\le 600$, while the small divisors are estimated by using
the classical Diophantine non-resonance condition for larger values of
the trigonometric degree $|\vet{k}|$. Due to the extremely small value
of the constant $\gamma$ entering in the non-resonance condition (see
the following statement~\ref{teo1.CAP}), the corresponding worsening
effect is quite evident.

In the case of the model considered in the present section, the
computer-assisted proof can be successfully completed by following the
approach detailed in~\cite{valloc2022} so as to apply the KAM theorem
in the version described in~\cite{steloc2012} at the very end of the
computational procedure. Our result can be summarized as follows.

\begin{theorem}{\bf{(Computer-assisted)}}
\label{teo1.CAP}
Consider the Hamiltonian $\Hscr^{(6)}\,$, reported in
formula~\eqref{Ham.initial.CAP}, or, equivalently, expanded as
in~\eqref{Ham.KAM.rbar} with $\bar{r}=6$ and truncated up to degree
$2$ in the actions and to trigonometrical degree $14$ in the
angles. Let us refer to the ball of radius $\rho$ centered in
$\vet{\omega}^{(*)}$, i.e.,
$$
\Sscr_{\rho}(\vet{\omega}^{(*)}) =
\{\vet{\omega} \in \reali^4 \, : \,
\|\vet{\omega}-\vet{\omega}^{(*)}\|_{\infty}\leq \rho \}\,,
$$
where
\begin{align*}
&\omega_1^{(*)}=-0.0061397976714045992\, ,& 
&\omega_2^{(*)}=-0.0034842628951575595 \, ,\\
&\omega_3^{(*)}=-0.0073217663368488322 \, , & 
&\omega_4^{(*)}=-0.0059275598828692194\, .
\end{align*}
For each $\vet{\omega}\in\Sscr_{\rho}(\vet{\omega}^{(*)})$
such that it satisfies the Diophantine condition 
$$
\big|\vet{k}\cdot\vet{\omega}\big|\geq
\frac{\gamma}{|\vet{k}|^{\tau}}
\qquad\forall\,\vet{k}\in\interi^{4}\setminus \{\vet{0}\}\, ,
$$
with\footnote{We remark that the subset of the Diophantine
  vectors with parameters $\gamma=2.9200551117155624\times 10^{-17}$
  and $\tau=4$ that are contained in the ball
  $\Sscr_{\rho}(\vet{\omega}^{(*)})$ is such that its Lebesgue measure
  is greater or equal to the 90$\%$ of the volume of the ball
  $\Sscr_{\rho}(\vet{\omega}^{(*)})$ itself, when $\rho=2\times
  10^{-15}$. Within the CAP such an estimate is done by applying a very
  standard argument about the measure of the resonant regions, which
  can be found, e.g., in appendix~A.2.1 of~\cite{gio2022}.}
$\rho=2\times 10^{-15}$, $\gamma= 2.9200551117155624\times 10^{-17} $
and $\tau=4\,$, there exists an analytic canonical transformation
leading the Hamiltonian $\Hscr^{(6)}$ in the Kolmogorov normal form
$\Hscr^{(\infty)}(\vet{P},\vet{Q})=\Escr^{(\infty)}+
\vet{\omega}\cdot\vet{P}+\Oscr(||\vet{P}||^2)\,$.  It is such
that, in the new variables, the torus $\{\vet{P}=\vet{0}\,,\,
\vet{Q}\in\toro^4\}$ is invariant and travelled by quasi-periodic
motions whose corresponding angular velocity vector is equal to
$\vet{\omega}$.
\end{theorem}

\section{KAM stability of the $2+3/2$ DOF secular model of the innermost  exoplanet in the \Ups system, by including also relativistic effects}
\label{sec:CAP_REL}

Recalling that the innermost planet of the \Ups system is very close
to a star that is about $30\%$ more massive than the Sun, one can
easily realize that corrections due to general relativity can play a
relevant role. As outlined in Section~$6$ of~\cite{masloc2023}, the
general relativity induces stabilizing effects on the orbital dynamics
of~\ups $\b$, producing also a significant modification of the
pericenter precession rate.  Similarly to what has been done in the
previous Section, also for a new model including relativistic
corrections, we want to rigorously prove the existence of KAM tori
(corresponding to suitably selected orbits) through a
computer-assisted proof.

\subsection{Secular orbital evolution of \ups $\b$ taking also into account relativistic effects}
\label{subsec:SQPR_REL}

The secular Hamiltonian, taking into account also general relativistic
effects, is defined so that
\begin{equation}
\label{mean_REL}
\Hscr^{(GR)}_{sec}=
\int_{\toro^3}\frac{\Hscr_{4BP}}{8\pi^3}\,  d\lambda_1 d\lambda_2 d\lambda_3 +
\int_{\toro} \frac{\Hscr_{GR}}{2\pi}\,  d M_1\,
:= \Hscr^{(NG)}_{sec}+\avg{\Hscr_{{GR}}}_{M_1} \, ,
\end{equation}
where $\Hscr_{4BP}$ defines the four body problem
(see~\eqref{Ham.4BP.reduced.mass}) and $\Hscr_{GR}$ describes the
general (post-Newtonian) relativistic corrections to the Newtonian
mechanics. More precisely, $\Hscr^{(NG)}_{sec}$ (recall
definition~\eqref{average_Ham4BP}) is explicitely written in
equation~\eqref{H4BP_expl}, while the average of the GR contribution
with respect to the mean anomaly $M_1$ of \ups$\b$ is such that
\begin{equation}
\label{Ham.sec.rel.qpr_1}
\avg{\Hscr_{{GR}}}_{M_1}=
-\frac{3\,\mathcal{G}^2\, m_0^2\, m_1}{ a_1^2 c^2\sqrt{1-\e_1^2}}
+\frac{15\,\mathcal{G}^2\, m_0^2\, m_1}{8 a_1^2 c^2}
-\frac{\mathcal{G}^2 \,m_0\, m_1^2}{8 a_1^2 c^2}\, ,
\end{equation} 
$c$ being the velocity of light in vacuum (see~\cite{miggoz2009} or
Appendix~E of~\cite{mas2023}).  Thus, the secular quasi-periodic
restricted model of the dynamics of \ups$\b$ which includes
corrections due to General Relativity (hereafter, SQPR-GR) can be
described by the following $2+3/2$~DOF Hamiltonian:
\begin{equation}
\begin{aligned}
\label{Ham.b.new_REL}
\Hscr^{(GR)}_{sec,\, 2+\frac{3}{2}}
&(\vet{p},\vet{q}, \csi_1, \eta_1, P_1, Q_1 )=
\omega_3\,p_3 + \omega_4\,p_4 + \omega_5\,p_5\\
&\qquad\phantom{=}+\Hscr^{(NG)}_{sec}(q_3,  q_4, q_5, \csi_1, \eta_1, P_1, Q_1 )
+\avg{\Hscr_{GR}}_{M_1}(\csi_1,\eta_1) \, ,
\end{aligned}
\end{equation}
where the angular velocity vector
$\vet{\omega}=(\omega_3,\omega_4,\omega_5\,)$ is given
in~\eqref{freq.fond.CD} and $\Hscr^{(NG)}_{sec}$ can be replaced by
$\Hscr^{1-2}_{sec}+\Hscr^{1-3}_{sec}$ appearing in
formula~\eqref{Ham.b.new}.  Finally, in the framework of this SQPR-GR
model, the equations for the orbital motion of the innermost planet
can be written as
\begin{equation}
\label{campo.Ham.b_REL}
\begin{cases}
  \dot{q_3}=\partial \Hscr^{(GR)}_{sec,2+\frac{3}{2}}/\partial p_3=\omega_3\\
  \dot{q_4}=\partial \Hscr^{(GR)}_{sec,2+\frac{3}{2}}/\partial p_4=\omega_4\\
  \dot{q_5}=\partial \Hscr^{(GR)}_{sec,2+\frac{3}{2}}/\partial p_5=\omega_5\\
  \dot{\csi}_{1}=
  -\partial \left(\Hscr_{sec}^{(NG)}(q_3, q_4, q_5, \csi_1, \eta_1, P_1, Q_1 )
  +\avg{\Hscr_{GR}}_{M_1}(\csi_1, \eta_1 )\right) /\partial \eta_1 \\
  \dot{\eta}_{1}=
  \partial \left(\Hscr_{sec}^{(NG)}(q_3, q_4, q_5, \csi_1, \eta_1, P_1, Q_1 )
  +\avg{\Hscr_{GR}}_{M_1}(\csi_1, \eta_1)\right) /\partial \csi_1 \\
  \dot{P}_{1}=
  -\partial \Hscr_{sec}^{(NG)}(q_3,  q_4, q_5, \csi_1, \eta_1, P_1, Q_1)
  /\partial Q_1 \\
  \dot{Q}_{1}=
  \partial \Hscr_{sec}^{(NG)}(q_3,  q_4, q_5, \csi_1, \eta_1, P_1, Q_1 )
  /\partial P_1
\end{cases}\, .
\end{equation}
Thus, it is possible to numerically integrate the equations of
motion~\eqref{campo.Ham.b_REL} of the SQPR-GR model in a similar way
to what has been done for the ones already described in
Section~\ref{sec:SQPR}.  Finally, let us remark that the invariance
law expressed in formula~\eqref{invariance} is fulfilled also with
$\Hscr^{(GR)}_{sec,\, 2+3/2}$, i.e., it holds true also for the
Hamiltonian term $\avg{\Hscr_{{GR}}}_{M_1}$ which takes into account
the relativistic effects on the orbital dynamics of \ups $\b\,$.

\subsection{$2+2/2$ DOF Hamiltonian model for the dynamics of \ups $\b$ with relativistic corrections}
\label{subsec:2+2/2DOF_REL}

Proceeding analogously to Sections~\ref{sec:TE}, we start from the
SQPR-GR Hamiltonian~\eqref{Ham.b.new_REL} and we perform the change of
variables~\eqref{coord:I_alpha}, arriving at the following compact
form of the Hamiltonian (similar to formula~\eqref{Ham_init_our}):
\begin{equation}
\label{Ham_init_our_REL}
\begin{aligned}
  \Hscr_{GR}^{(0)}(\vet{p},\vet{q}, \vet{I},\vet{\alpha})=
  \Hscr^{(GR)}_{sec,\, 2+3/2}(\vet{p},\vet{q},\vet{I},\vet{\alpha})
  &=
  \Escr_{GR}^{(0)}+\vet{\omega}^{(0)}\cdot\vet{p}+
  \vet{\Omega}_{GR}^{(0)}\cdot\vet{I}+
  \sum_{s= 0}^{\mathcal{N}_S}\sum_{l= 3}^{\mathcal{N}_L}
  h_{l}^{(0,s)}(\vet{q},\vet{I},\vet{\alpha})\\
  &\phantom{=}+\sum_{s= 1}^{\mathcal{N}_S}\sum_{l=0}^{2}
  h_{l}^{(0,s)}(\vet{q},\vet{I},\vet{\alpha}) \, ,
\end{aligned}
\end{equation}
with $\Escr_{GR}^{(0)}$ constant,
$(\vet{\omega}^{(0)},\vet{\Omega}_{GR}^{(0)})\in\reali^3\times
\reali^2$ the angular velocity vector\footnote{The value of
  $\vet{\omega}^{(0)}$ is reported in formula~\eqref{freq.fond.CD} and
  it is not affected by the relativistic corrections, being the
  angular velocity vector related to the external exoplanets.} and
$h_{l}^{(0,s)}\in\mathfrak{P}_{l,sK}$. In agreement with
footnote~\ref{foot_param}, we take $\mathcal{N}_L=6$ as maximal power
degree in the square root of the actions and we include Fourier terms
up to a maximal trigonometric degree of $10$, putting
$\mathcal{N}_S=5$ and $K=2\,$. Thus, we perform $\mathcal{N}_S$ steps
of the algorithm constructing the normal form for an elliptic torus
(according to Section~\ref{sec:TE}), so as to introduce
$$
\Hscr_{GR}^{(\mathcal{N}_S)}(\vet{p},\vet{q}, \vet{I},\vet{\alpha})
=\Escr_{GR}^{(\mathcal{N}_S)}+\vet{\omega}\cdot\vet{p}
+\vet{\Omega}_{GR}^{(\mathcal{N}_S)}\cdot\vet{I}
+\sum_{s=0}^{\mathcal{N}_S}\sum_{l= 3}^{\mathcal{N}_L}
h_{l}^{(\mathcal{N}_S,s)}(\vet{q},\vet{I},\vet{\alpha})\, ,
$$
which is very similar to the Hamiltonian defined
in~\eqref{Ham.T.E.r.final.S}.  Moreover, proceeding as in
Section~\ref{sub:appl_KAM_2+3/2degree}, we can perform the change of
variables~\eqref{change_loose_1DOF}, arriving at a Hamiltonian with
the same form as that written in formula~\eqref{Ham.Kam.moreDOF.iniz},
i.e.,
\begin{equation}
\label{Ham.Kam.moreDOF.iniz_REL}
\Hscr_{ GR, K}^{(0)}(\vet{P},\vet{Q};\,\vet{I}_\ast) =
\Escr_{GR}^{(0)}(\vet{I}_\ast)+\big(\vet{\omega}_{GR}^{(0)}(\vet{I}_\ast)\big)\cdot\vet{P}
+\sum_{s=0}^{\t{\mathcal{N}}_S}\sum_{l= 2}^{\t{\mathcal{N}}_L}
h_{l}^{(0,s)}(\vet{P},\vet{Q};\vet{I}_\ast)
+\sum_{s=1}^{\t{\mathcal{N}}_S}\sum_{l= 0}^{1}
h_{l}^{(0,s)}(\vet{P},\vet{Q};\vet{I}_\ast)\, ,
\end{equation}
where $(\vet{P},\vet{Q}):= (P_1,P_2,P_3,P_4,Q_1,Q_2,Q_3,Q_4)\,$, and
$\vet{\omega}_{GR}^{(0)}(\vet{I}_\ast)\in\reali^4$ is defined so that
\begin{equation}
\label{new_freq_lose_REL}
\begin{aligned}
  &\omega_{GR,\,1}^{(0)}(\vet{I}_\ast)=
  \frac{\partial\avg{\Hscr_{GR, K}^{(0)}}_{\vet{Q}}}{\partial P_1}
  \Bigg|_{\substack{P_1=0\\ P_2=0}}\,, &
  &\omega_{GR, \,2}^{(0)}(\vet{I}_\ast)=
  \frac{\partial\avg{\Hscr_{GR,K}^{(0)}}_{\vet{Q}}}{\partial P_2}
  \Bigg|_{\substack{P_1=0\\ P_2=0}}\,,\\
  &\omega_{GR,3}^{(0)}=\omega_3^{(0)}=\omega_3-\omega_5\,, &
  &\omega_{GR,4}^{(0)}=\omega_4^{(0)}=\omega_4-\omega_5\,.
\end{aligned}
\end{equation}
Finally, in order to construct the Kolmogorov normal form, we proceed
analogously to Section~\ref{subsec:results_KAM_no_freq_fixed}; thus,
we start from
$\Kscr^{(0)}(\vet{P},\vet{Q})=\Hscr_{GR,K}^{(0)}(\vet{P},\vet{Q};\,\vet{I}_\ast)$,
expressed in formula~\eqref{Ham.Kam.moreDOF.iniz_REL} and we apply the
Kolmogorov normalization algorithm (see
Section~\ref{subsec:KAM_no_freq_fixed}) in junction with a Newton-like
method, performing $\bar{r}=\t{\mathcal{N}}_S\,$ normalization steps
by using {\tt Mathematica} as an algebraic manipulator. Thus, we
arrive at the following Hamiltonian that is truncated and in
Kolmogorov normal form:
\begin{align}
\label{Ham.KAM.fine.2+3/2DOF_REL}
 \Hscr_{GR,K}^{(\t{\mathcal{{N}}}_S)}(\vet{P},\vet{Q};\,\vet{I}_{\ast}^{(\nNewt)}) =
 \Escr_{GR}^{(\t{\mathcal{{N}}}_S)}(\vet{I}_{\ast}^{(\nNewt)})
 +\big(\vet{\omega}_{GR}^{(\t{\mathcal{{N}}}_S)}(\vet{I}_{\ast}^{(\nNewt)})\big)
 \cdot\vet{P}
 +\sum_{s=0}^{\t{\mathcal{N}}_S}\sum_{l= 2}^{\t{\mathcal{N}}_L}
 h_{l}^{(\t{\mathcal{{N}}}_S,s)}(\vet{P},\vet{Q};\,\vet{I}_{\ast}^{(\nNewt)})\, ,
\end{align}
which is analogous to the one written in
formula~\eqref{Ham.KAM.fine.2+3/2DOF}.
In order to find $\vet{I}_{\ast}^{(\nNewt)}$, we iterate the Newton
method until
$$
\big\|\Delta\vet{\omega}^{(GR)}(\vet{I}_{\ast}^{(\nNewt)})\big\|_{\infty} =
\big\|\big(\omega_{GR,1}^{(\t{\mathcal{N}}_S)}({\vet{I}}_{\ast}^{(\nNewt)})-
\omega_{GR,1}^{(*)}\,,\,
\omega_{GR,2}^{(\t{\mathcal{N}}_S)}({\vet{I}}_{\ast}^{(\nNewt)})-
\omega_{GR,2}^{(*)}\big) \big\|_{\infty}<10^{-10}\, ,
$$
where the values of the components of the angular velocity vector
$\vet{\omega}_{GR}^{(*)}\,$ are such that
\begin{equation}
\label{freq_punto_REL}
\omega_{GR,1}^{(*)}=\omega_{GR,1}-\omega_5\,,
\quad
\omega_{GR,2}^{(*)}=\omega_{GR,2}-\omega_5\,,
\quad
\omega_{GR,3}^{(*)}=\omega_3-\omega_5\,,
\quad
\omega_{GR,4}^{(*)}=\omega_4-\omega_5\,.
\end{equation}
In the previous formula, the values of
$(\omega_3,\omega_4,\omega_5)\in\reali^3$, are related to the
fundamental periods of the two outer exoplanets and are given in
equation~\eqref{freq.fond.CD}, while $\omega_{GR,1}$ and
$\omega_{GR,2}$ are the values of the fundamental angular velocities
as obtained by the frequency analysis method when it is applied to the
discretized signals
$t\mapsto\sqrt{2\Lambda_1}\,\sqrt{1-\sqrt{1-\big(\e_1(t)\big)^2}}
\,e^{-i\varpi_1(t)}\,$, $t\mapsto
2\sqrt{\Lambda_1}\,\sqrt[4]{1-\big(\e_1(t)\big)^2}
\,\sin\big(\frac{\i_1(t)}{2}\big)\,e^{-i\Omega_1(t)}$, respectively,
which are produced by the numerical integration of the original
$2+3/2$ DOF Hamiltonian model with GR corrections, defined in
formula~\eqref{Ham.b.new_REL}.

\subsection{Results produced by the semi-analytic integration of the SQPR Hamiltonian model with $2+2/2$ DOF and GR corrections}
\label{subsub:risut_KAM_REL}

We now start considering the numerical integration of secular
quasi-periodic restricted Hamiltonian which has $2+3/2$~DOF and takes
into account the relativistic effects, namely $\Hscr^{(GR)}_{sec,\,
  2+\frac{3}{2}}$ written in~\eqref{Ham.b.new_REL}, whose
corresponding equations of motion are reported in
formula~\eqref{campo.Ham.b_REL}. We consider the initial conditions
which are reported in Table~\ref{tab:param.orb.b_GR} and approximately
correspond to the center of the stability region according to the
numerical study described in Section~6 of~\cite{masloc2023}.

\begin{table}[h]
\begin{minipage}{0.5\textwidth}
\begin{center}
\begin{tabular}{ll}
\toprule
 & $\upsilon$-And $\bf{b}$  \\
\midrule
$m \, [M_J]$ & $0.674$ \\
$a(0) \, [{\rm AU}]$ & $0.0594$  \\
$\e(0)$ & $0.011769$ \\
$\i(0)\, [^{\circ}]$ & $20.$ \\
$M(0)\, [^{\circ}]$ & $103.53$ \\
$\omega(0)\, [^{\circ}]$ & $51.14$ \\
$\Omega(0)\, [^{\circ}]$ & $0.$ \\
\bottomrule
\end{tabular}
\end{center}
\end{minipage}
\begin{minipage}{0.5\textwidth}
\caption{Values of the initial orbital parameters for \ups$\b$. The
  chosen value of the mass is the minimal one according
  to~\cite{mcaetal2010}. The values $a_1(0)\,$, $\e_1(0)\,$, $M_1(0)$
  and $\omega_1(0)$ are reported from the stable prograde trial
  PRO2 of~\cite{deietal2015} (Table~$3\,$). The values of initial
  inclination and longitude of the node (denoted with $\i_1(0)$ and
  $\Omega_1(0)$, respectively) are taken from~\cite{masloc2023}; see
  the text for more details. The corresponding initial orbital
  parameters in the Laplace reference frame can be easily determined
  (see, e.g.,~\cite{mas2023}).}
\label{tab:param.orb.b_GR}
\end{minipage}
\end{table}

\begin{figure}[h]
\begin{minipage}{.5\textwidth}
\includegraphics[scale=0.36]{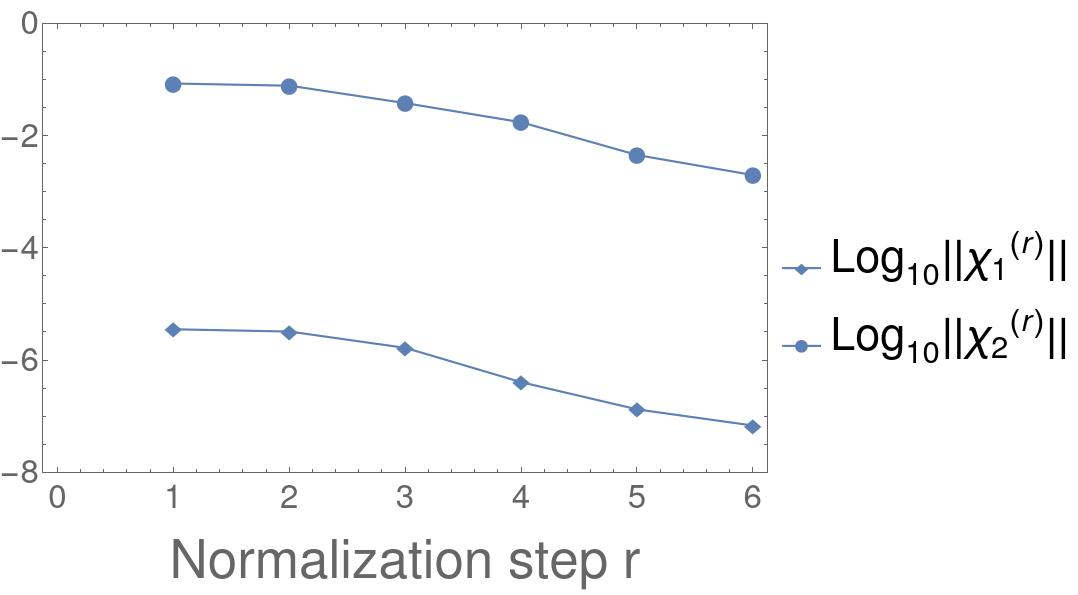}
\end{minipage}
\qquad\qquad\qquad\qquad
\begin{minipage}{.28\textwidth}
\caption{Convergence of the generating functions $\chi_1^{(r)}$ and
  $\chi_2^{(r)}$ defined by the normalization algorithm {\`a la}
  Kolmogorov without keeping fixed the angular velocity vector, when
  it is applied to the 2+2/2~DOF SQPR model {\it with} GR
  corrections and is performed in the case corresponding to $(\i_1(0),
  \Omega_1(0))=(20^{\circ},0^{\circ})$. The ${\rm Log}_{10}$ of their
  norms are reported as a function of the normalization step $r\,$.}
\label{fig.norm_decay_2+3/2_REL}
\end{minipage}
\end{figure}

For what concerns the semi-analytical approach, we start from the
$2+2/2$DOF Hamiltonian, written
in formula~\eqref{Ham.KAM.fine.2+3/2DOF_REL}. For the construction of the
Kolmogorov normal form (without fixing the angular velocity vector, as
explained in Section~\ref{subsec:KAM_no_freq_fixed} and applied in the
previous Section), we adopt $\t{\mathcal{N}}_ L = 2\,$, $\t{K}= 2\,$,
$\t{\mathcal{N}}_S = 6\,$ as parameters ruling the truncations of the
expansions. In particular, $\nNewt=2$ iterations of the Newton method
are enough to reach the condition
$\|\vet{\omega}_{GR}^{(\t{\mathcal{{N}}}_S)}({\vet{I}}_{\ast}^{(2)})
-\vet{\omega}_{GR}^{(*)}\|_{\infty}<10^{-10}$, which allows us to
successfully conclude the search for the initial translation vector
that approximates well enough the one we are looking for, namely the
unknown ${\vet{I}}_{\ast}\,$.  We focus again on the last Kolmogorov
normalization that is performed at the end of the Newton method, i.e.,
the one corresponding to the initial translation vector
${\vet{I}}_{\ast}^{(2)}$. Looking at
Figure~\ref{fig.norm_decay_2+3/2_REL}, once again we can appreciate a
rather regular and sharp decay of the norms of the generating
functions $\chi_1^{(r)}$ and $\chi_2^{(r)}$ .

\begin{figure}[h]
\begin{minipage}{.45\textwidth}
\includegraphics[scale=0.28]{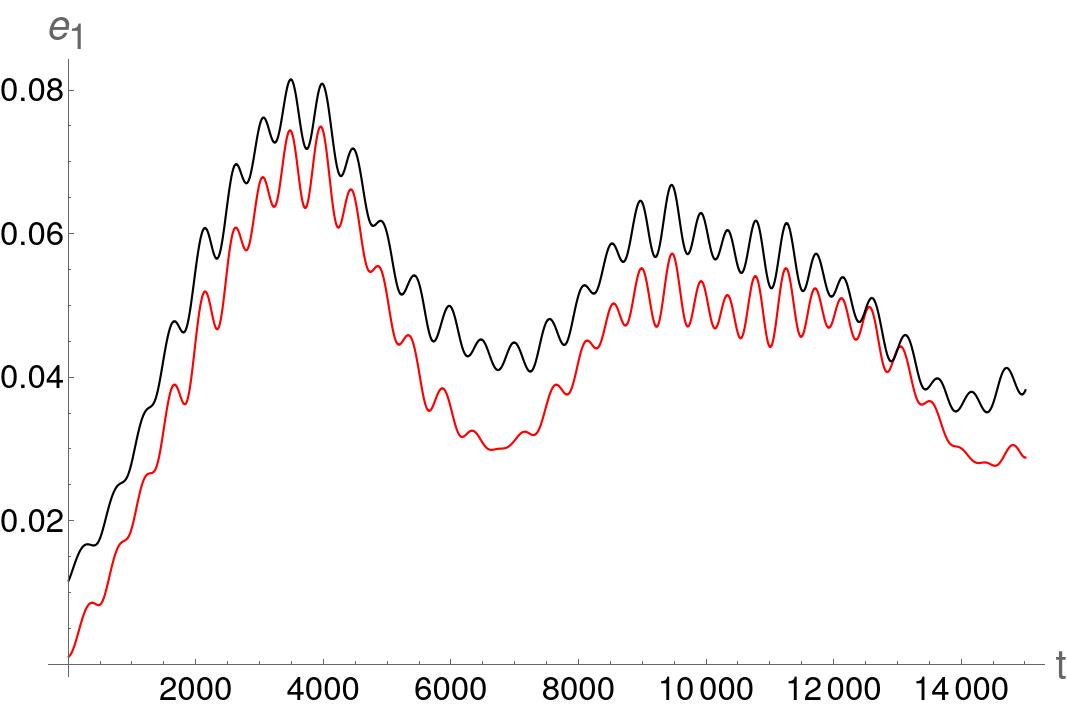}
\end{minipage}
\quad\quad
\begin{minipage}{.45\textwidth}
\includegraphics[scale=0.28]{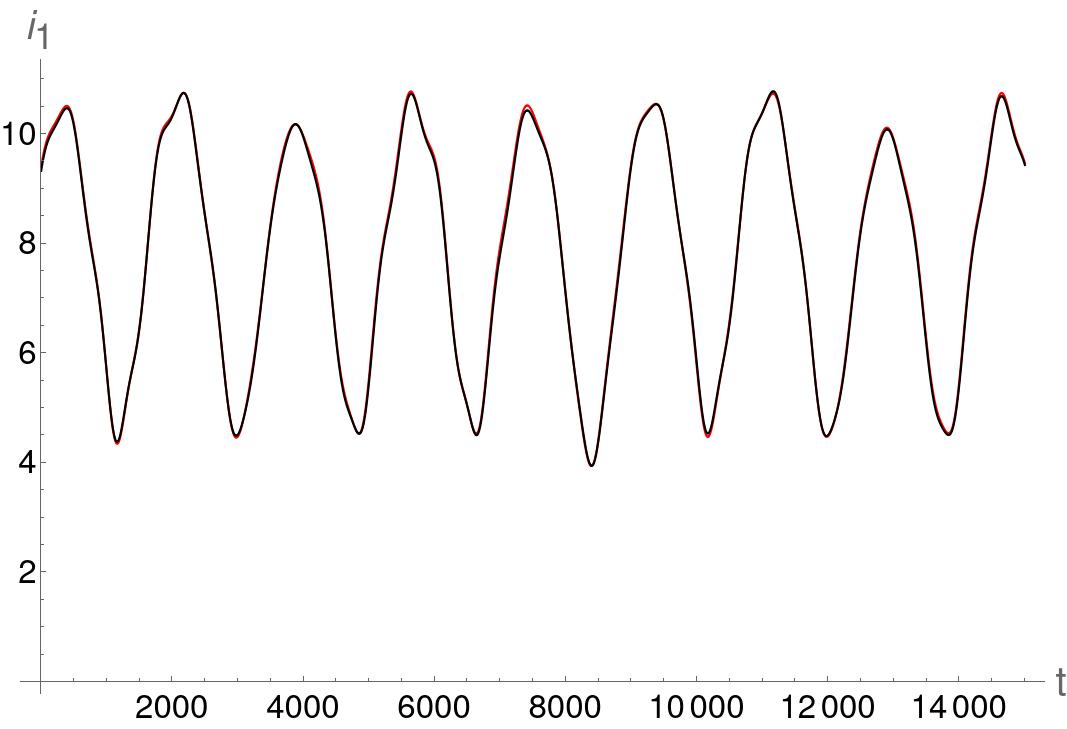}
\end{minipage}
\caption{Comparisons between the eccentricity $\e_1$ (on the left) and
  the inclination $\i_1$ (on the right) as obtained through the
  semi-analytical approach (in red) and the numerical one (in black).
  Both the integration methods consider the case with initial
  conditions referring to the data reported in
  Table~\ref{tab:param.orb.b_GR} and they are applied to $2+3/2$~DOF
  model which {\it does} take into account the effects due to GR.
  Units of measure for what concerns inclination $\i_1$ and time $t$
  are degree and year, respectively.}
\label{fig.comparison_e1_i1_2+3/2_REL}
\end{figure}

Following again the approach described in the previous
Subsection~\ref{subsub:risut_KAM}, we can construct a semi-analytic
(approximate) solution of the equations of
motion~\eqref{campo.Ham.b_REL} which here corresponds to a
quasi-periodic orbit whose angular velocity vector is now denoted with
$\vet{\omega}_{GR}^{(*)}\,$. Such a quasi-periodic evolution is
plotted (in red) in Figure~\ref{fig.comparison_e1_i1_2+3/2_REL}, where
one can appreciate that there is rather good agreement with the
numerical integrations of the motion (in black) for what concerns
the behavior of both the eccentricity and the inclination of \ups$\b$
(apart a small shift of the two plots reported in the left panel).

\subsection{Computer-Assisted Proofs of existence of KAM tori for the SQPR Hamiltonian models with $2+2/2$ DOF and GR corrections}
\label{subsec:CAP_REL}

In a similar way to what has been done in Section~\ref{sec:CAP}, we
want to adopt an approach based on rigorous CAPs also to the case
described above, i.e., adding also the relativistic effects on the
dynamics of \ups $\b$.

Thus, we use the publicly available software
  package${{\ref{footnote-CAP4KAM_nDOF}}\atop{\phantom{1}}}$ that is
  designed for performing CAPs, in order to apply it to the new
initial Hamiltonian
\begin{equation}
\label{Ham.initial.CAP_REL}
\begin{aligned}
  &H_{GR}^{(0)}(\vet{P},\vet{Q})=
  \Hscr_{GR}^{(\t{\mathcal{{N}}}_S-1)}(\vet{P},\vet{Q};\,\vet{I}_{\ast}^{({2})}) 
  \\
  &= \Escr_{GR}^{(\t{\mathcal{{N}}}_S-1)}(\vet{I}_{\ast}^{({2})})
  +\big(\vet{\omega}_{GR}^{(\t{\mathcal{{N}}}_S-1)}(\vet{I}_{\ast}^{({2})})\big)\cdot
  \vet{P} +\sum_{s=0}^{\t{\mathcal{N}}_s}\sum_{l= 2}^{\t{\mathcal{N}}_L}
  h_{l}^{(\t{\mathcal{{N}}}_S-1,\,s)}(\vet{P},\vet{Q};
  \,\vet{I}_{\ast}^{({2})})+\sum_{l= 0}^{1}
  h_{l}^{(\t{\mathcal{{N}}}_S-1,\,
    \t{\mathcal{N}}_S)}(\vet{P},\vet{Q};\,\vet{I}_{\ast}^{({2})})\,,
\end{aligned}
\end{equation}
with $\t{N}_L=2$, $\t{K}=2$ and $\t{N}_S=7\,$.\footnote{In this case we
  have put $\t{N}_S=7$ so that, as explained in the previous
  Subsection,
  $\vet{\omega}_{GR}^{(6)}({\vet{I}}_{\ast}^{({2})})\simeq\vet{\omega}_{GR}^{(*)}$,
  since $\big\|\big(
  \omega_{GR,1}^{(6)}({\vet{I}}_{\ast}^{({2})})-\vet{\omega}_{GR,1}^{(*)}\,,\,
  \omega_{GR,2}^{(6)}({\vet{I}}_{\ast}^{({2})})-\vet{\omega}_{GR,2}^{(*)}\big)
  \big\|_{\infty}<10^{-10}$. } As remarked in Section~\ref{sec:CAP},
performing also the last step would completely remove all the
perturbing terms that are represented in our {\it truncated}
expansions, making the application of a CAP to
$\Hscr_{GR}^{(\t{\mathcal{{N}}}_S)}$ meaningless.

In the case of the $2+2/2$ DOF Hamiltonian model
$\Hscr^{(\t{\mathcal{{N}}}_S-1)}(\vet{P},\vet{Q};\,\vet{I}_{\ast}^{({2})})$
(which consider also relativistic effects, described in
formula~\eqref{Ham.initial.CAP_REL}) the CAP succeeds in rigorously
proving the existence of a set (with positive Lebesgue measure) of KAM
tori whose corresponding angular velocity vectors are Diophantine and
in an extremely small neighborhood of $\vet{\omega}_{GR}^{(*)}\,$.  In
the present case the values of the parameters that are internal to the
CAP and that mostly affect the computational complexity are fixed
so that $R_{\rm{I}}=38$ and $R_{\rm{II}}=54000$. The CPU-time needed
to complete the CAP is about $32.2$~days on a workstation equipped with
processors of type \texttt{Intel XEON-GOLD 5220} (2.2~GHz). It is
amazing that our CAP in the present case needs just $\simeq 43\%$ of
CPU-time with respect to the application described in
Section~\ref{sec:CAP}. This is mainly due to the fact that the
generating functions $\chi_1^{(r)}$ and $\chi_2^{(r)}$ (when they are
explicitly computed for $1\le r\le R_{\rm{I}}$) contain less terms in
their Fourier expansions and their norms decrease more sharply. Also
this phenomenon can be seen as a byproduct of the stabilizing effect
due to the relativistic corrections.

\begin{figure}[!h]
\begin{minipage}{.48\textwidth}
\includegraphics[scale=0.68]{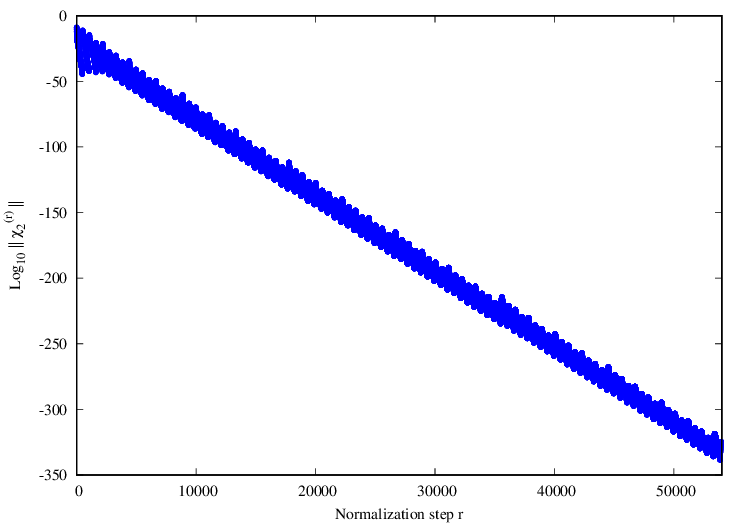}
\end{minipage}
\quad
\begin{minipage}{.48\textwidth}
\includegraphics[scale=0.68]{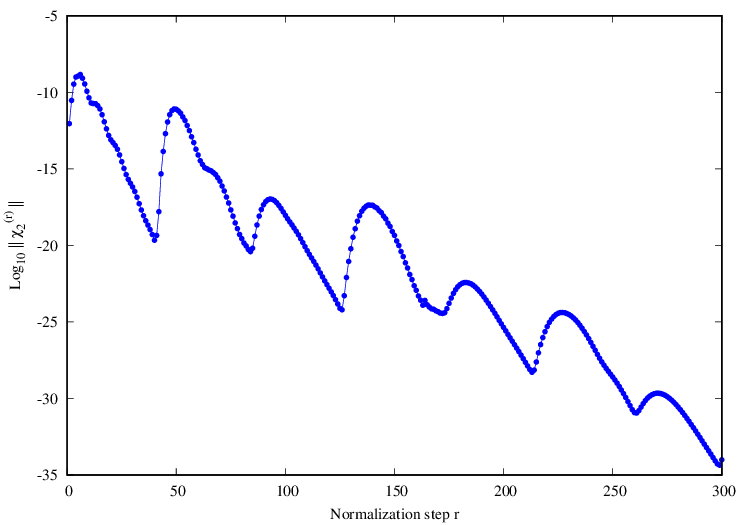}
\end{minipage}
\caption{Estimates of the norms (in semi-log scale) of the generating
  function $\chi_2^{(r)}$, as produced during the CAP, using
  $R_{\rm{I}}=38$ and $R_{\rm{II}}=54000$. On the right, the zoom on
  the first $300$ normalization steps.}
\label{fig.CAP_REL}
\end{figure}

The plot (in semi-log scale) of the estimates of the norm of
$\chi_2^{(r)}$, reported in Fig.~\ref{fig.CAP_REL}, shows a regular
decreasing behaviour. Moreover, looking at the plot on the right, it
is evident that such a decrease is steeper up to the step
$R_{\rm{I}}$. As in Fig.~\ref{fig.CAP}, both in the left panel of
Fig.~\ref{fig.CAP_REL} and in the one on the right, we can appreciate
the periodicity of small jumps, that depend on the parameters {\tt
  MAXMODKCALC} (here fixed so as to be equal to $550$) and
$R_{\rm{I}}\,$, respectively; both of these parameters are internal to
the CAP.

Also in this case case, the computer-assisted proof can be
successfully completed by applying the KAM theorem in the version
described in~\cite{steloc2012}. Our final result is summarized by the
following statement.

\begin{theorem}{\bf{(Computer-assisted)}}
\label{teo2.CAP}
Consider the Hamiltonian $\Hscr_{GR}^{(6)}\,$, reported in
formula~\eqref{Ham.initial.CAP_REL}. Let us introduce the ball of
radius $\rho$ centered in $\vet{\omega}_{GR}^{(*)}$, i.e.,
$$
\Sscr_{\rho}(\vet{\omega}_{GR}^{(*)}) =
\{\vet{\omega} \in \reali^4 \, : \,
\|\vet{\omega}-\vet{\omega}_{GR}^{(*)}\|_{\infty}\leq \rho \}\,,
$$
where
\begin{align*}
&\omega_{GR,1}^{(*)}=-0.0064644895081070399\, ,& 
&\omega_{GR,2}^{(*)}=-0.0034914960690528801 \, ,\\
&\omega_{GR,3}^{(*)}=-0.0073217663368488322 \, , & 
&\omega_{GR,4}^{(*)}=-0.0059275598828692194\, .
\end{align*}
For each $\vet{\omega}\in\Sscr_{\rho}(\vet{\omega}_{GR}^{(*)})$
such that it satisfies the Diophantine condition 
$$
\big|\vet{k}\cdot\vet{\omega}\big|\geq
\frac{\gamma}{|\vet{k}|^{\tau}}
\qquad\forall\,\vet{k}\in\interi^{4}\setminus \{\vet{0}\}\, ,
$$ with $\rho=2\times 10^{-15}$, $\gamma= 2.6761115506846878\times
10^{-17} $ and $\tau=4\,$, there exists an analytic canonical
transformation leading $\Hscr_{GR}^{(6)}$ in a Kolmogorov normal form
for which there is a torus which is invariant with respect to the
corresponding Hamiltonian flow and is travelled by quasi-periodic
motions whose corresponding angular velocity vector is equal to
$\vet{\omega}$.
\end{theorem}

\section*{Acknowledgments}
We are very grateful to Prof.~Christos Efthymiopoulos for his
encouragements and suggestions.  This work has been partially
supported by the MIUR Excellence Department Project MatMod@TOV
(2023-2027) awarded to the Department of Mathematics, University of
Rome ``Tor Vergata'' and by the Spoke 1 ``FutureHPC \& BigData''
  of the Italian Research Center on High-Performance Computing, Big
  Data and Quantum Computing (ICSC) funded by MUR Missione 4
  Componente 2 Investimento 1.4: Potenziamento strutture di ricerca e
  creazione di ``campioni nazionali'' di R{\&}S (M4C2-19) - Next
  Generation EU (NGEU).

\bibliographystyle{abbrv}
%\nocite{*}
\bibliography{biblio}

\end{document}